\documentclass{aa}

\usepackage{txfonts}

\usepackage[english]{babel}
\usepackage{amsmath,amssymb}
\usepackage{graphicx, subfig}
\usepackage[normalem]{ulem}

\usepackage{verbatim}

\usepackage{multirow}
\usepackage{mathtools}
\usepackage{epstopdf}

\usepackage{url}
\usepackage{xcolor,color}

\usepackage{soul}
\usepackage{orcidlink}

\usepackage{natbib,twoopt}
\usepackage[hyphenbreaks]{breakurl}

\usepackage{tabularx}

\usepackage{placeins}

\bibpunct{(}{)}{;}{a}{}{,}             
\definecolor{cobalt}{rgb}{0.06, 0.2, 0.65}
\hypersetup{
  colorlinks,
  citecolor=cobalt,
  linkcolor=[rgb]{0.8, 0.2, 1.0},
  urlcolor=cobalt,
}
\makeatletter
  \newcommandtwoopt{\citeads}[3][][]{\href{http://adsabs.harvard.edu/abs/#3}%
    {\def\hyper@linkstart##1##2{}%
     \let\hyper@linkend\@empty\citealp[#1][#2]{#3}}}
  \newcommandtwoopt{\citepads}[3][][]{\href{http://adsabs.harvard.edu/abs/#3}%
    {\def\hyper@linkstart##1##2{}%
     \let\hyper@linkend\@empty\citep[#1][#2]{#3}}}
  \newcommandtwoopt{\citetads}[3][][]{\href{http://adsabs.harvard.edu/abs/#3}%
    {\def\hyper@linkstart##1##2{}%
     \let\hyper@linkend\@empty\citet[#1][#2]{#3}}}
  \newcommandtwoopt{\citeyearads}[3][][]%
    {\href{http://adsabs.harvard.edu/abs/#3}
    {\def\hyper@linkstart##1##2{}%
     \let\hyper@linkend\@empty\citeyear[#1][#2]{#3}}}
\makeatother

\newcommand{\unit}[1]{\, {\rm #1}}

\newcommand\code[1]{\textsc{#1}}

\newcommand{\quotes}[1]{``#1''}

\newcommand{\commenta}[1]{#1}

\def\apjl{ApJL}
\def\apjs{ApJS}
\def\aap{A\&A}

\begin{document}

\title{Stellar halos tracing the assembly of ultra-faint dwarf galaxies}
\titlerunning{Stellar halos around ultra-faint dwarf galaxies}

\author{
Lapo Querci \orcidlink{0009-0006-7675-2614} \inst{1}\fnmsep\thanks{\href{mailto:lapo.querci1@unifi.it}{lapo.querci1@unifi.it}} \and
Andrea Pallottini \orcidlink{0000-0002-7129-5761} \inst{2} \and
Lorenzo Branca \orcidlink{0000-0002-6064-1964} \inst{2,3} \and
Stefania Salvadori \orcidlink{0000-0001-7298-2478} \inst{1,4}.}
\authorrunning{Querci et al.}

\institute{
Dipartimento di Fisica e Astronomia, Università degli Studi di Firenze, Via G. Sansone 1, 50019, Sesto Fiorentino, Italy \and
Scuola Normale Superiore, Piazza dei Cavalieri 7, 56126 Pisa, Italy \and
Interdisciplinary Center for Scientific Computing, University of Heidelberg, Im Neuenheimer Feld 205, D-69120 Heidelberg, Germany \and
INAF - Osservatorio Astrofisico di Arcetri, Largo E.Fermi 5, 50125, Firenze, Italy 
}

\date{Received 3 October, 2024/ Accepted 18 December, 2024}

\abstract
{Ultra-faint dwarfs (UFDs) are expected to be relics of the earliest galaxies to have formed in the Universe. Observations show the presence of a stellar halo around UFDs, which can give precious insights into UFD evolution. Indeed, stellar halos can form via tidal interaction, early supernova feedback, or merging events.
}
{This work investigates how merger properties impact the formation of stellar halos around UFDs, focusing on Tucana II, the most promising UFD assembled through mergers.}
{We developed N-body simulations of dry isolated mergers between two UFDs, resolving their stellar component down to $1 \unit{M_\odot}$.
We built a suite of simulations by varying:  the merger-specific i) angular momentum, $l$, and ii) kinetic energy $k$, iii) the merger mass ratio, $M_1/M_2$, iv) the dark-to-stellar mass ratio, $M_{\rm DM}/M_\star$, of the progenitors, and v) their stellar size, $R_{1/2}$.
To fully explore such a five-dimensional parameter space, we trained a neural network to emulate the properties of the resulting \quotes{post-merger} UFD, by quantifying the half-mass radius ($R_\star$) and the fraction of stars at radii $>5R_\star$ ($f_5$). 
}
{Our principal component analysis clearly shows that $f_5$ ($R_\star$) is primarily determined by $M_1/M_2$ ($R_{1/2}$), with $R_{1/2}$ ($M_1/M_2$) playing a secondary role. Both $f_5$ and $R_\star$ show almost no dependence on $k$, $l$, and $M_{\rm DM}/M_\star$ in the explored range.
Using our emulator, we find that to form the stellar halo observed in Tucana~II; that is, $f_5=10\pm5\%$ and $R_\star=120\pm 30 \unit{pc}$, we need to merge progenitors with $M_1/M_2 = 8_{-3}^{+4}$, the size of the more massive one being $R_{1/2} = 97^{+25}_{-18} \unit{pc}$.
Such findings are corroborated by the consistency ($\chi^2\simeq 0.5 -2$) between stellar density profiles observed for Tucana~II and those of simulations that have $M_1/M_2$ and $R_{1/2}$ close to the values predicted by the emulator.
}
{The stellar halos of UFDs contain crucial information about the properties of their smaller progenitor galaxies. Ongoing and planned spectroscopic surveys will greatly increase the statistics of observed stars in UFDs, and thus of their associated stellar halos. By interpreting such observations with our simulations, we will provide new insights into the assembly history of UFDs, and thus the early galaxy formation process.
}

\keywords{galaxies: dwarf, evolution, interactions, Local Group -- methods: numerical}

\maketitle



\section{Introduction}\label{sec:intro}

Dwarf galaxies have been extensively observed in the Local Group, where more than 70 objects have been detected as satellites orbiting around the Milky Way (MW) and Andromeda \citep{McConnachie:2012}. Among these satellites, the ultra-faint dwarf galaxies \citep[UFDs; $\rm{M} \lesssim 10^5 \unit{M_\odot}$,][]{simon:2019} are dominated by ancient, $>10 \unit{Gyr}$ \citep{Brown:2012, Gallart:2021}, and very metal-poor stars \citep{Kirby:2008, Spite:2018} and, similarly to the classical dwarf spheroidal galaxies, have little to no gas mass \citep[e.g.,][]{Westmeier:2015}. 
The observed properties of UFDs, which include the metallicity distribution function and the fraction of carbon-enhanced metal-poor stars \citep[CEMP-no, e.g.,][]{Lai:2011, Yoon:2019} strongly support the idea that UFDs are living relics of the first star-forming mini-halos \citep[e.g.,][]{Ricotti:2005, salvadori:2009, Jeon:2017, barnes_hierarchical_1986}, which formed before the end of reionization and are hosted by \commenta{$\sim 10^6-10^8 \unit{M_\odot}$ dark matter halos.}
Their early formation and low mass make them the best candidates to host the first stars' descendants \citep[e.g.,][]{Salvadori:2015, Magg:2018, Rossi:2024} and perfect laboratories in which to study the effects of feedback processes on galaxy formation \citep[e.g.,][]{Jeon:2015,agertz:2020,gutcke:2022}.
Moreover, UFDs are the most dark-matter-dominated stellar systems known, with mass-to-luminosity ratios of $M/L > 100 \unit{M_\odot/L_\odot}$, and studying them is expected to play a key role in helping us understand the nature of dark matter \citep[e.g.,][]{Bullock:2017, Strigari:2018, battaglia:2022}.

Identifying UFD stars becomes increasingly difficult as the distance from the galactic center increases due to their faint nature and the foreground contamination of MW stars.
Nonetheless, recent observations of UFDs have started to provide evidence for stars at large distances from the center. This is the case for Tucana~II, where stars have been identified up to a half-light radius of $\sim 9$ $R_{1/2}$ \citep{chiti:2021}, Hercules~I, with stars detected up to $\sim 9.5\, R_{1/2}$ \citep{Garling:2018,Longeard:2023, Ou:2024}, Tucana~V, up to $\sim 10\, R_{1/2}$ \citep{Hansen:2024}, Ursa Major~I up to $\sim 4\, R_{1/2}$, Bo\"otes~I  up to $\sim 4 R_{1/2}$, and Coma Berenice up to $\sim 2\, R_{1/2}$ \citep{Waller:2023}.
In addition, multiple studies identify new and distant stars exploiting Gaia data. \citet{Tau:2024} explore the outskirts of 30 UFDs using already known and new RR~Lyrae stars and show that at least $\sim30\%$ present extended stellar populations. By considering $\sim \, 60$ MW dwarf galaxies' satellites, \citet{Jensen:2024} systematically identify candidate member stars based on spatial, color-magnitude, and proper motion information; they report an extended secondary outer profile for nine dwarf MW satellites, seven of which are UFDs: Bo\"otes I, Bo\"otes III, Draco II, Grus II, Segue I, Tucana II, and Tucana III. 

The formation channels for such extended features can be i) tidal stripping, ii) in situ formation via supernova (SN) feedback, and iii) merging events.
In proximity to the MW, its gravitational pull can deform the UFD's morphology, forming an extended halo, as is shown by \citet{Fattahi:2018}. Signatures of tidal interaction are signaled by the presence of velocity gradients \citep{Li:2018} and deviations from the expected exponential profile that typically describes the stellar density distribution \citep{Penarrubia:2008}.
Early SN feedback can induce an inside-out stellar formation, with the successive generation of stars forming at larger radii. This scenario is plausible, since UFDs may have a bursty star formation history \citep{Wheeler:2019} -- that is, high rates of star formation lasting a short period of time -- as is expected for galaxies that formed most of their stars prior to reionization \citep{Gelli:2020, Gelli:2023, pallottini:2023, sun:2023}. \commenta{However, compared to more massive galaxies, the star formation rate (SFR) of UFDs is expected to be low -- $< 10^{-4} \unit{M_\odot yr^{-1}}$ \citep{Salvadori:2014} -- and thus the stellar feedback scenario requires energetic SNe to compensate for the low SN rate.} Signs of SNe can be retrieved in the chemical composition of stars, and thus high-resolution spectroscopy is the primary way to determine if the UFD stellar halos have an in situ origin \citep[e.g.,][]{chiti:2023}.      
Merger events between UFDs are expected \citep{Deason:2014, Revaz:2023} and they can form extended stellar components beyond  $5 - 20\, R_{1/2}$ \citep{deason:2022, Ricotti:2022}. Determining if a stellar halo formed via merging events requires a combination of spatial and chemical information to rule out the other scenarios. Currently, only in Tucana~II does the stellar halo seem consistent at a high confidence level with the merger scenario \citep{chiti:2021, chiti:2023}.

Tucana~II is a UFD with a stellar mass of $M_\star^{\rm TucII} = 3^{+7}_{-2} \times 10^3 \unit{M_\odot}$ and a half-light radius of $R_{1/2}^{\rm TucII} = 120\pm 30 \unit{pc}$ \citep{Bechtol:2015}, and it has one of the lowest average metallicities in the Local Group, $<[\rm{Fe}/\rm{H}]> = -2.7$ \citep{Ji:2016, Chiti:2018}. 
When observing the outskirts of Tucana~II, \citet{chiti:2021} identified five and two giant stars located at distances of $>2\, R_{1/2}$ and $> 5\, R_{1/2}$, respectively.
\cite{chiti:2021, chiti:2023} rule out the tidal interaction scenario for Tucana II because of: i) the orthogonality of the stellar extension with respect to the predicted direction of the tidal debris; and ii) the absence of a velocity gradient in the extended stellar component. \commenta{The position of the pericenter \citep[around $\sim 40\, \unit{kpc}$;][]{battaglia:2022, Pace:2022} is distant enough to corroborate a scenario of low impact from tidal interactions, even though recent studies of cosmological zoom-in simulations suggest that satellites experience tidal disruption at comparable pericenters \citep[e.g.][]{Shipp:2023, Shipp:2024}.}
The SN feedback scenario for the formation of the stellar halo in Tucana~II is disfavored because of the low SFR and for two additional reasons. First, \citet{chiti:2023} report that 75\% of the observed metal-poor stars ([Fe/H]< -2.9) are carbon-enhanced (i.e., the so-called CEMP-no, [C/Fe]>0.7), which suggests enrichment by low-energy primordial “faint SNe” \citep[e.g.,][]{Salvadori:2015, Jeon:2017, Rossi:2023}. Second, the chemical abundances of stars in Tucana~II at various [Fe/H] are consistent with the ones observed in other UFDs. Therefore, if SN feedback is the main formation channel of stellar halos, most UFDs should present an extended feature, which instead is not observed.
Ruling out the other possibilities, the remaining formation scenario for Tucana~II stellar halo is the merging of two UFDs.

A result supporting such a scenario comes from the cosmological hydrodynamical simulations by \citet{Tarumi:2021}. They report that a galaxy closely resembling Tucana~II in terms of the fraction of distant stars, metallicity, and metallicity gradient can be formed by a major merger; that is, a merger for which the mass ratio between the two galaxies is within 1:1 and 10:1.
%
There have been other studies analyzing the impact of dwarf galaxy mergers on the formation of stellar halos. Using N-body simulations, \citet{deason:2022} find that intermediate merger ratios ($\sim5$:1) maximize the growth of extended stellar halos regardless of the stellar mass–halo mass relation. By preparing cosmological N-body simulations, \citet{Ricotti:2022} show that the shape and mass of the stellar halos can be described as the sum of nearly exponential profiles, with scale radii determined by the mass ratio of the mergers building it. \citet{Goater:2024} analyzed tidally isolated UFDs in zoom-in hydrodynamical simulations, taken from the EDGE cosmological simulation suite \citep{agertz:2020}, finding that many galaxies exhibit anisotropic extended stellar halos mimicking tidal tails, but such halos are actually formed by late-time mergers.
These studies primarily focus on the influence of the merger mass ratio on the stellar halo formation, with limited exploration of other potential merger or progenitor characteristics, leaving open questions such as whether the merger trajectory has any impact on the stellar halo formation, whether the dark matter total mass influences the size of the post-merger galaxy or the stellar halo formation, and what properties of the merger we can infer using present-day observations. 

In this work, we explore the parameter space of UFD mergers, considering five key properties -- the specific angular momentum, specific kinetic energy, merger mass ratio, total UFD mass, and initial radius -- to model the trajectory and mass ratio of the merger and the mass and radius of the progenitors.
Our goal is to investigate how these parameters influence the formation of the stellar halo in UFDs. We developed a suite of N-body simulations of isolated mergers between two possible Tucana~II progenitors, to perform a wide (>1 dex) parameter space exploration.
\commenta{Ideally, we would need: i) a large number of simulations, to perform a high-frequency sampling of the parameter space, and ii) single-stellar mass resolution, to have good precision in the identification of merger debris.} To achieve both features, we developed an emulator that can efficiently generate the properties of the post-merger galaxies, once trained with simulation data.
To validate our findings, we conducted a comparative analysis of simulated stellar surface density profiles and the one observed in Tucana~II.

This article is structured as follows. First, we describe the simulation setup, the parameter space explored, and how we constructed the emulator (Sect. \ref{sec:methods}). Then, the results of the parameter space exploration and the comparison with observations are presented (Sect. \ref{sec:results}). Finally, we summarize and discuss our findings in (Sect. \ref{sec:summary_and_discussion}).


\section{Methods}\label{sec:methods}

Here, we present the suite of simulations for isolated merging UFD galaxies and detail our strategy for exploring the parameter space of the possible merger configurations.

We start by reviewing the base assumptions and the schemes used to generate the initial conditions (ICs) for the progenitors (Sect. \ref{sec:sub:initial_conditions}).
Then, we discuss the intervals adopted for the ICs (Sect. \ref{sec:sub:parameter_space}) and show how we can explore them with a Latin hypercube sampling of the merger parameter space (Sect. \ref{sec:sub:parameter_space_sampling}).
Finally, we describe the numerical code used to follow the time evolution of the merger (Sect. \ref{sec:sub:numerical_code}) and we explain how we trained an emulator by using the results from our suite (Sect. \ref{sec:sub:merger_emulator}).

\subsection{Initial conditions}\label{sec:sub:initial_conditions}

Our merging simulations were set up at redshift $z = 6$, after the reionization of the Universe, when most of the gas of UFDs -- that is, minihalos -- is expected to be removed or heated by radiation \citep{barkana:1999,sobacchi:2013,gutcke:2022}. For this reason, we neglected the gas component and focused on dry mergers; that is, mergers of galaxies with only dark matter and stars.

Throughout the simulation suite, we varied the main properties of the dark matter and stellar component of the two merging galaxies together with the parameters defining the kinematic of the merger.
Specifically, we explored: i) the dark matter mass, $M_{\rm DM}$; ii) the stellar mass, $M_\star$; iii) the scale radius of the progenitors; and iv) the trajectory of the merger. A detailed discussion of the ranges adopted for these quantities is given in Sect. \ref{sec:sub:parameter_space} (see Table \ref{table:parameterSpace_intervals} for a summary), while, in the following, we describe the general setup and parametrization of the ICs.

\subsubsection{Dark matter profile}

We assume the dark matter density profile, $\rho_{\rm DM}$, to be a spherically symmetric Navarro-Frank-White \citep{navarro:1996}:
\begin{align}\label{eq:NFW}
    \rho_{\rm DM}(r) &= \frac{3 M_{\rm DM}}{4 \pi R_{vir}^3} \left( \ln(1+c) - \frac{c}{1+c}\right)^{-1}\times \\
    & \times \frac{r}{R_{vir}}^{-1}\left(\frac{1}{c} +\frac{r}{R_{vir}}\right)^{-2} \,, \nonumber
\end{align}
where $r$ is the distance from the galaxy center, $R_{vir}$ is the virial radius, and $c$ is the concentration parameter. \commenta{Considering the virial radius, defined as the radius where the density is $\Delta_c =200$ times the critical density of the Universe, we can express it as:
\begin{equation}
    R_{vir} = \left[\frac{(M_{\rm DM} + M_\star) G}{100\, H^2}\right]^{1/3} \,,
\end{equation} 
with $G$ being the gravitational constant and $H=H(z)$ the Hubble parameter}\footnote{The cosmological parameters adopted in this work are taken from \citet{Planck:2015}: $H_0 = 67.74 \unit{km}\unit{s^{-1}}\unit{Mpc^{-1}}$, $\Omega_m = 0.3089$, and $\Omega_\Lambda = 0.6911$.}; that is,
\begin{equation}
    H = H_0\left[ \Omega_m (1+z)^3 + \Omega_\Lambda\right]^{1/2}\,.
\end{equation}
The concentration is $c = 3.5$, assuming the results found by \citet{diemer:2015} for galaxies hosted in $M_{\rm DM}=10^8 \unit{M_\odot}$ halos at $z = 6$.

\subsubsection{Stellar profile}
We adopted an exponential profile, which is commonly utilized to fit stellar density profiles of UFDs \citep[e.g.,][]{Belokurov:2006, Martin:2008, Munoz:2018}. The profile is defined by
\begin{equation}\label{eq:exponentialProfile}
    \rho_\star(r) = \rho_{0} e^{-r/(0.5R_{1/2})}\,,
\end{equation}
where $\rho_{0}$ is the central density, $R_{1/2}$ is the projected radius that contains half of the mass of the galaxy, in other words the half-mass radius, and the $0.5$ factor arises integrating the exponential profile, factor $0.351$, and projecting the three-dimensional scale radius to the projected two-dimensional $R_{1/2}$, factor $1.33$ \citep{wolf:2010}. In the suite, we varied $R_{1/2}$ to account for different initial stellar densities, as is discussed in Sect. \ref{sec:sub:parameter_space}. 

\subsubsection{Discretization of the progenitors}\label{sec:dice}

\begin{figure}
  \includegraphics[width=0.49\textwidth]{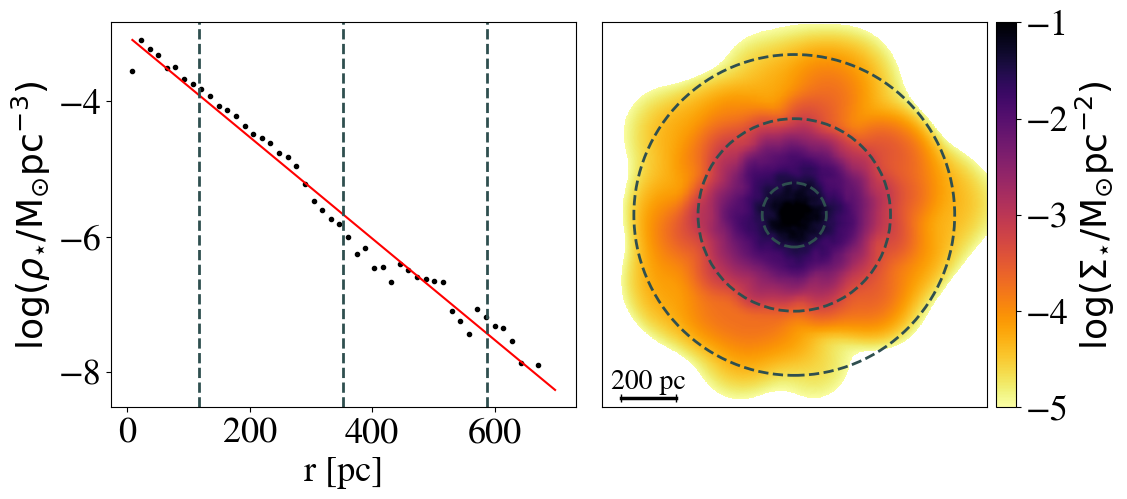}
  \caption{
  Example of the virialized stellar distribution of a progenitor with $M_\star = 5.48 \times10^3 \unit{M_\odot}$ and $R_{1/2} = 118\unit{pc}$ generated using \code{DICE} (see Sect. \ref{sec:dice}).
  Left panel: Stellar density ($\rho_\star$) extracted by spherically averaging the mass in a simulation shell (points) and the fit exponential profile  (solid line, Eq. \ref{eq:exponentialProfile}) as a function of the radius ($r$).
  Right panel: Stellar surface density map, $\Sigma_\star$ integrated along the line of sight; the spatial scale is indicated as an inset.
  In both panels, the dashed lines mark the location of spheres with radius $R_{1/2}$, $3\,R_{1/2}$, and $5\,R_{1/2}$.
  \label{fig:progenitor_IC}
  }
\end{figure}

We used the public code \code{DICE} \citep{perret_dice_2016} to perform a Lagrangian sampling of the particles composing each progenitor in order to obtain the selected density profiles for dark matter (Eq. \ref{eq:NFW}) and stars (Eq. \ref{eq:exponentialProfile}).
\code{DICE} uses an N-try Markov chain Monte Carlo method to extract the particles' position, $\vec{x}_s$, from the profiles and integrates the Jeans equations to assign the velocity, $\vec{v}_s$, for each particle, $s$; this is done separately for the progenitors.

To match the observed stellar mass of Tucana~II, we fixed the sum of the stellar mass of the two progenitors to $M_{\star, tot} = 6000 \unit{M_\odot}$.
To compare our results with star-by-star observations, we opted for a single-stellar mass resolution; in all the simulations, the mass resolution of the stellar component is $m_{\star} = 1 \unit{M_\odot}$.
The number of dark matter particles was adjusted to have a mass resolution of $m_{\rm DM} = 100 \unit{M_\odot}$. The dark matter to stellar mass resolution ratio adopted is consistent with the one used in \citet{Ricotti:2022} and satisfies the trade-off between computational cost and spurious numerical heating \citep[e.g.,][]{Ludlow:2023, DiCintio:2024}. The numerical heating mostly affects the size of the post-merger galaxy, while it is negligible for the stellar halo formation, as is shown in Appendix \ref{sec:app:resolution_DM}.
%

As an example of the procedure, in Fig. \ref{fig:progenitor_IC} we report the stellar density profile (left panel) and the stellar surface density map on the $XY$ plane (right panel) generated by \code{DICE} for a progenitor with $M_\star = 5.48 \times 10^3 \unit{M_\odot}$ and $R_{1/2} = 118\, \unit{pc}$. In the left panel, the solid red line shows the exponential profile fit to the progenitor density profile. 
Since \code{DICE} computes galaxies at Jeans' equilibrium, the initial progenitors are not at virial equilibrium and must evolve in isolation to virialize. We address this aspect in the next paragraph (Sect. \ref{sec:merger_setup}).

\subsubsection{Merger setup}\label{sec:merger_setup}

\begin{figure}
  \includegraphics[width=0.49\textwidth]{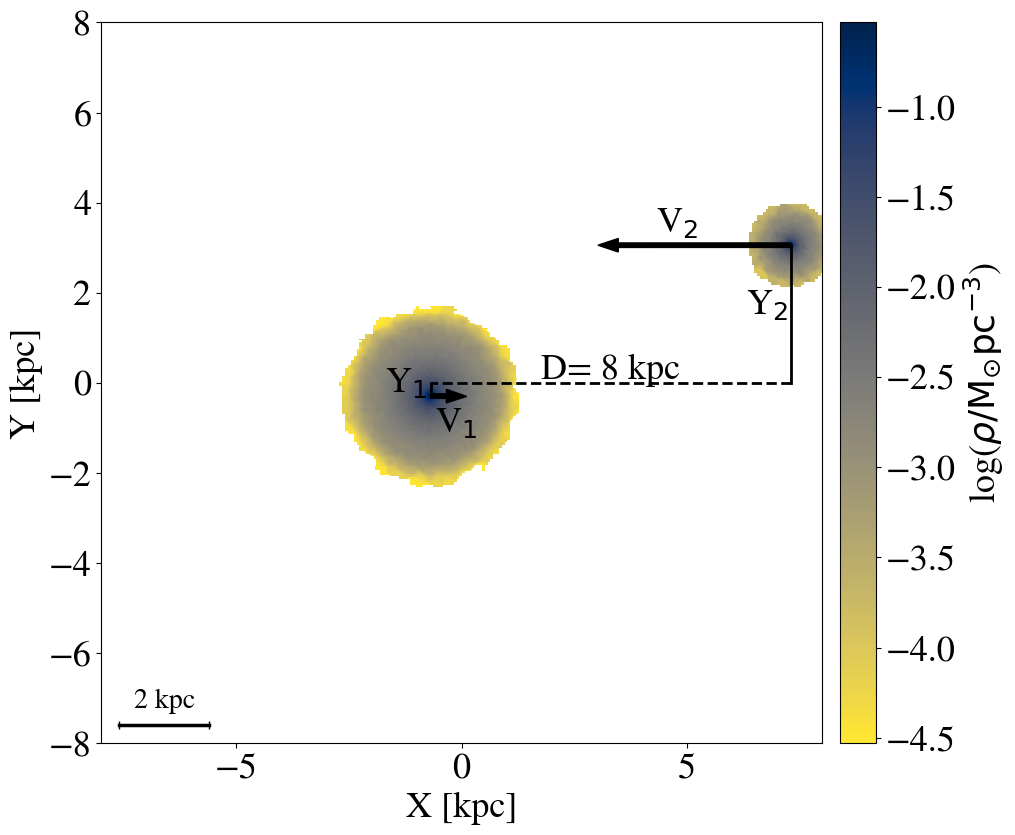}
  \caption{
  Example of merger ICs prepared with \code{DICE} (see Sect. \ref{sec:merger_setup}).
  The density map in the X-Y plane shows the total (dark matter+stars) density ($\rho$) that has been mass-averaged along the Z axis.
  The two progenitors were placed in the simulation box at a distance of $D=8\,\rm kpc$ along the X axis and with an impact parameter of $|Y_1| + |Y_2|$ (Eq. \ref{eq:def:y_position_simulationBox}).
  The arrows mark the direction of the initial velocity of each progenitor (Eq. \ref{eq:def:relative_velocity}).
  \label{fig:merger_IC}
  }
\end{figure}

Using \code{DICE}, we placed the two progenitors in the simulation box, assigned them a relative velocity, set the system's center of mass at rest in the center, and placed the progenitors at a distance of $D = 8 \unit{kpc}$ along the X axis, which ensured that they virialized while the gravitational interaction was still negligible.
Hereafter, the subscript 1 refers to the more massive progenitor and 2 to the other one.

We set the progenitors on the same Z axis, assuming that the velocity differs from zero only along the X axis. The initial position of the progenitors can be written as
\begin{subequations}\label{eq:IC_posizioni_e_velocita}
\begin{align}
    X_{1,2} &= \mp D\frac{M_{2,1}}{M_1 + M_2}\,,\\
    Y_{1,2} &= \mp \frac{l}{|V_1| + |V_2|}\frac{M_1+M_2}{M_{1,2}}\,,
    \label{eq:def:y_position_simulationBox}
\end{align}
with their velocity along the X axis being
\begin{equation}\label{eq:def:relative_velocity}
    V_{1,2}  = \pm \sqrt{2k\frac{M_{2,1}}{M_{1,2}}}\,.
\end{equation}
\end{subequations}
In Eq. \ref{eq:IC_posizioni_e_velocita}, $M_i$ is the total mass of the $i$-th progenitor ($M_i = M_{i, \rm DM} + M_{i, \star}$), $l$ is the specific angular momentum (with regard to the center of mass),
\begin{subequations}
\begin{equation}\label{eq:def:specific_L}
l = \frac{\sum_s  m_s \vec{v}_s \times \vec{x}_s}{\sum_s m_s}\,,
\end{equation}
where the sum over $s$ is extended to all particles, and $k$ is the specific kinetic energy,
\begin{equation}\label{eq:def:specific_K}
k = \frac{\sum_s  m_s \vec{v}_s^2 }{\sum_s m_s}\,.
\end{equation}
\end{subequations}

We varied the values of $l$ and $k$ to probe different merging trajectories.
Figure \ref{fig:merger_IC} shows an example of the merger's ICs, in which the two progenitors have total masses of $M_{1} = 4.8\times 10^7 \unit{M_\odot}$ and $M_{2} = 4.6\times 10^6 \unit{M_\odot}$, the specific merger kinetic energy is $k =0.72 \unit{pc}^2\unit{Myr}^{-2}$, and the angular momentum is $l=1.1\times10^3 \unit{pc}^2\unit{Myr}^{-1}$.

\subsection{Parameterization of the merger}\label{sec:sub:parameter_space}

\begin{table}
    \caption{Parameters intervals adopted in the suite
    }
    \label{table:parameterSpace_intervals}
    \centering
    \begin{tabular}{l  c c l l}
        \hline                        
        parameter   & min & max & unit  & definition\\
        \hline                                     
        $M_1/M_2$             & $1$             &$30$             & -                             & Sect. \ref{sec:merger_mass_ratio} \\      
        $M_{\rm DM}/M_\star$  & $8\times 10^3 $ &$ 10^5$          & -                             & Sect. \ref{sec:dm_to_star}\\
        $R_{1/2}$             & $15 $           &$ 150 $          & $\unit{pc}$                   & Sect. \ref{sec:half_mass_radius}, Eq. \ref{eq:exponentialProfile}\\          
        $l$                   & $ 10^2 $        &$ 3\times 10^3 $ & $\frac{\unit{pc}^2}{\unit{Myr}}$  & Sect. \ref{sec:k_and_l}, Eq. \ref{eq:def:specific_L} \\      
        $k$                   & $ 0.1 $         &$3 $             & $\frac{\unit{pc}^2}{\unit{Myr}^2}$ & Sect. \ref{sec:k_and_l}, Eq. \ref{eq:def:specific_K}\\
        \hline    
    \end{tabular}
\end{table}

We focus our exploration on the five parameters: the merger mass ratio, $M_1/M_2$, the dark matter content, $M_{\rm DM}/M_{\star}$, the stellar radius of the progenitors, $R_{1/2}$, the specific angular momentum, $l$, and the specific kinetic energy, $k$.
In the following, we first describe the interval explored for each of these properties and then present how we explore the parameter space via a Latin hypercube sampling.
Table \ref{table:parameterSpace_intervals} summarizes the properties and relative intervals explored.  

\subsubsection{Merger mass ratio}\label{sec:merger_mass_ratio}

Defining the merger mass ratio between the two progenitors as $M_1/M_2$, we note that previous works \citep[e.g.,][]{deason:2022} identify a peak in the fraction of stars in the stellar halo when $M_1/M_2 \simeq 5$.
Therefore, to fully capture its dependency in UFD mergers, we explored the interval $M_1/M_2 \in [1,30]$.

It is convenient to use the merger mass ratio to define the stellar mass of each progenitor: 
\begin{subequations}
\begin{align}
    M_{1,\star} &= M_{tot,\star}\frac{M_1/M_2}{1+M_1/M_2}\\
    M_{2,\star} &= M_{tot,\star}\frac{1}{1+M_1/M_2}\, , 
\end{align}
\end{subequations}
where $M_{tot,\star}= M_{1,\star} + M_{2,\star}$ is the total stellar mass in the simulation, which was set to $M_{tot,\star} = 6 \times 10^3 \unit{M_\odot}$ to ensure that, even in the case of stellar mass loss during the merger, the post-merging galaxy stellar mass is consistent with Tucana~II estimations, $M_\star^{\rm 
TucII} = 3^{+7}_{-2} \times 10^3 \unit{M_\odot}$ \citep{Bechtol:2015}.

\subsubsection{Dark matter to stellar mass ratio}\label{sec:dm_to_star}

\begin{figure}
  \includegraphics[width=0.45\textwidth]{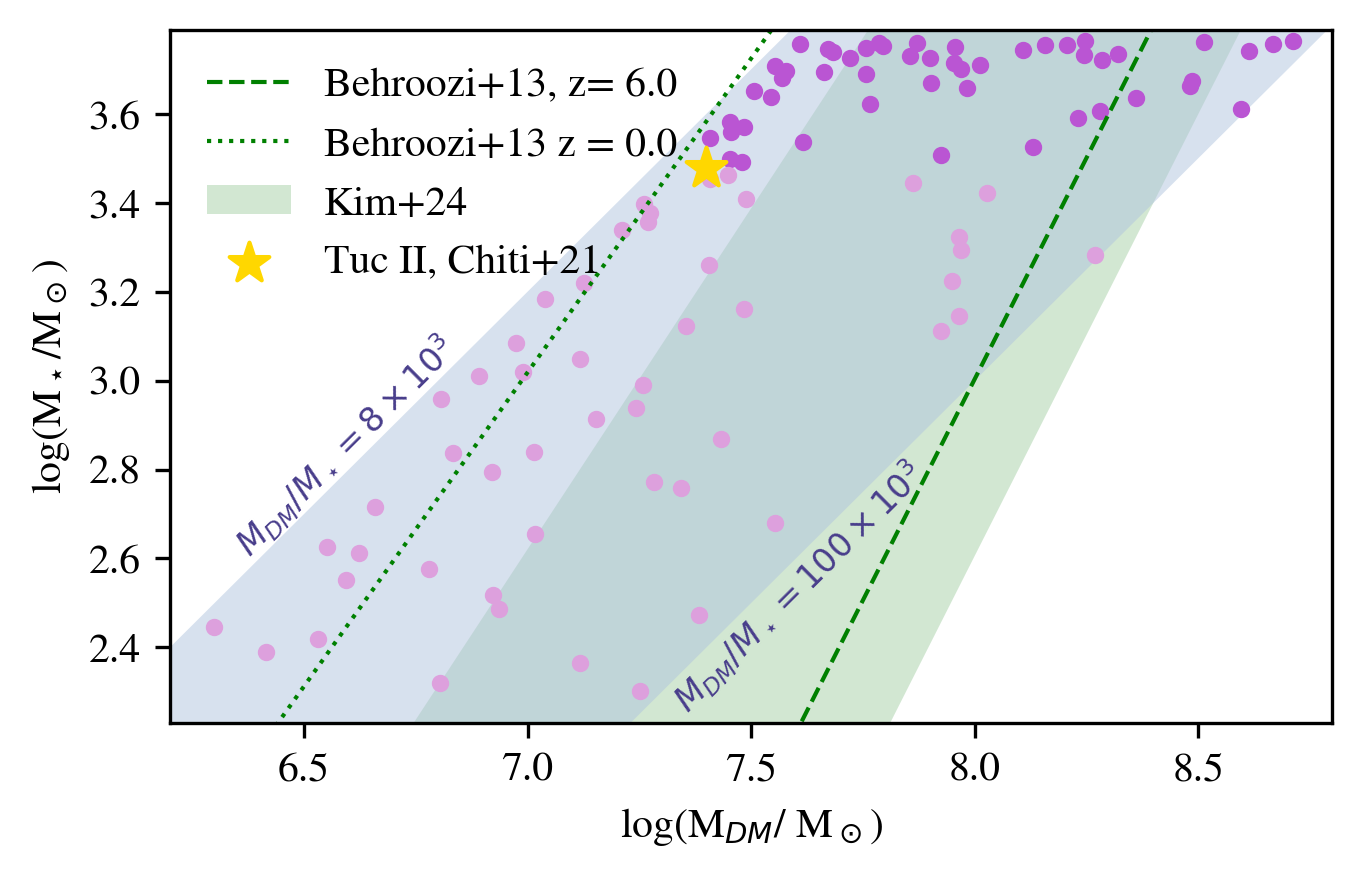}
  \caption{Parameter sampling in the dark matter ($M_{\rm DM}$) vs. stellar mass ($M_\star$) plane.
  Each pink (purple) point represents the least (more) massive progenitor in a merging simulation.
  The shaded blue region indicates the dark matter to stellar mass ($M_{\rm DM}/M_\star$) interval explored by the suite.
  The gold star shows the estimate of Tucana~II from \citet{chiti:2021} observations, while the green lines are the extrapolation from the abundance matching model from \citet{behroozi:2013} at redshift $z = 0$ and $6$. \commenta{The shaded green region represents the extrapolation of the scatter in the stellar-mass to halo-mass relation measured by \cite{Kim:2024} for dwarf galaxies.} 
  \label{fig:merger_mass_ratio_exploration}
  } 
\end{figure}

The dark matter content of UFDs is poorly constrained. Velocity dispersion observations can determine the dark matter mass within the luminous part of the galaxy; however, extrapolating the total halo masses from such a measurement is an uncertain endeavor \citep[e.g.,][]{Errani:2018}.
For this reason, we explored an interval of values of $M_{\rm DM}$, investigating whether the dark matter abundance impacts the stellar halo formation. To this end, we varied the stellar to dark matter mass ratio, $M_{\rm DM}/M_\star$, which determines the dark matter mass of each progenitor once the merger mass ratio sets the stellar mass.

The interval explored is $M_{\rm DM}/M_\star \in [8,100] \times 10^3$, and, in Fig. \ref{fig:merger_mass_ratio_exploration}, we report it in an $M_\star$ - $M_{\rm DM}$ plane with a shaded region.
The selected range includes the latest dark matter mass estimate within $1 \unit{kpc}$ for Tucana~II, $M_{\rm DM}^{\rm TucII} \sim 2.5 \times 10^7 \unit{M_\odot}$ \citep{chiti:2021}.
Further, the explored range roughly brackets the predictions by abundance-matching methods performed by \citep[][]{behroozi:2013} for galaxies between $z=6$ and $z=0$ (green lines in Fig. \ref{fig:merger_mass_ratio_exploration}). \commenta{Recently, a substantial effort has been carried out to constraint the stellar mass for smaller system both from local galaxy observations \citep[e.g.,][]{jethwa:2018, Nadler:2020} and simulations \citep[e.g.,][]{Munshi:2021, Kim:2024}. The explored range in this work encompasses most of the scatter measured at $z=0$ in simulated dwarf galaxies by \citet[][green-shaded area in Fig.\ref{fig:merger_mass_ratio_exploration}]{Kim:2024}}.

\subsubsection{Half-mass radius}\label{sec:half_mass_radius}

\begin{figure}
  \includegraphics[width=0.45\textwidth]{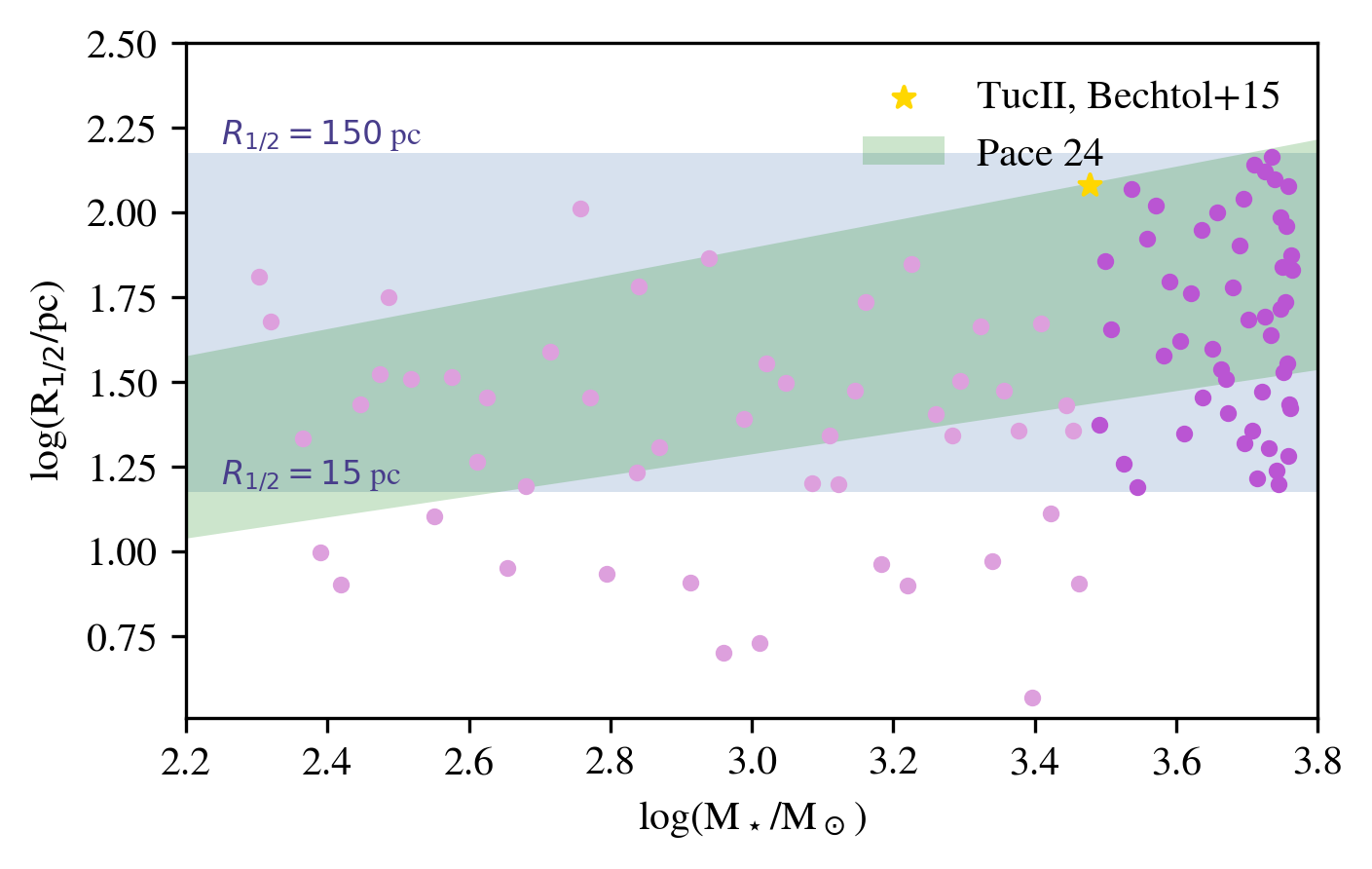}
  \caption{ \commenta{Parameter sampling in the stellar mass ($M_\star$) vs. scale radius ($R_{1/2}$) plane.
  Each pink (purple) point represents the least (more) massive progenitor in a merging simulation and the shaded blue area indicates the scale radius interval explored in the suite.
  The shaded green region represents the polynomial fit $\pm 1 \sigma$ of all the available UFD data in the Local Volume Database \citep{Pace:2024} .} 
  \label{fig:mass_radius_exploration}
  } 
\end{figure}

Observations show that UFDs with a stellar mass similar to Tucana~II have a half-light radius -- the radius that contains half of the galaxy's luminosity -- ranging from tens to hundreds of parsecs \citep{simon:2019}.
The half-light radius is equal to the half-mass radius, assuming a constant luminosity-to-mass ratio, and thus we decided to explore an interval of values, $R_{1/2}\in [15,150] \unit{pc}$.

We assigned the value of $R_{1/2}$ to the more massive progenitor, and we defined the mean stellar density as
\begin{equation}
    \bar\rho_\star = \frac{3}{8\pi } \frac{M_{1,\star}}{R_{1/2}^3}\,.
\end{equation}
To set the scale radius of the other progenitor, we imposed that the two galaxies have the same $\bar\rho_\star$; that is, the scale radius of the less massive was determined by $R_{1/2}(M_2/M_1)^{1/3}$. This approach prevented us from exploring the dependency of the stellar halo formation on the density difference between the two progenitors. However, by avoiding double sampling the $R_{1/2}$ parameter, we reduced the number of dimensions explored, thereby improving the robustness of our results.

\commenta{In Fig. \ref{fig:mass_radius_exploration}, we report the sampling of the $M_\star$-$R_{1/2}$. We considered the stellar mass and half-light radii of all the UFDs in the Local Universe Database \citep{Pace:2024} and performed a fit with a polynomial function. The fit $\pm1-\sigma$ range is reported as the green-shaded area in Fig.  \ref{fig:mass_radius_exploration}. The overlap between the sampled values and the observations in the $M_\star$-$R_{1/2}$ plane confirms that the adopted scaling and range of the radii are adequate to explore the physical parameter space.}

\subsubsection{Specific angular momentum and kinetic energy}\label{sec:k_and_l}

Since we want to study the merging event between galaxies, we limited our explorations to values of $l$ and $k$ that are highly likely to produce the merging of the two progenitors. Specifically, the intervals explored are $l \in [100,3000] \unit{pc}^2 \unit{Myr}^{-1}$ and $k \in [0.1,3] \unit{pc}^2 \unit{Myr}^{-2}$. We adopted these intervals after testing them with (i) the solution of the two-body problem, assuming both galaxies as point masses, and (ii) a low-resolution version of the simulation suite, as is further explained in the following. 

\subsection{Parameter space sampling}\label{sec:sub:parameter_space_sampling}

Sampling the parameter space is a delicate endeavor. The more samples we have, the more finely we explore the parameter space, with increasing computational cost.
We used a five-dimensional Latin hypercube sampling \citep{LHS} to maximize the exploration, while keeping the number of samples (simulations) low. The implemented sampling consists of 47 nodes (sampled points) with each node being a unique set of the five parameters defining the merger; that is,
\begin{equation}\label{eq:def:sampled_node}
    \boldsymbol\kappa = (l, k, M_1/M_2, M_{\rm DM}/M_\star, R_{1/2},)\,.
\end{equation}
To compute the node values with the LHS prescription, we divided the range of each individual parameter into a set of equally probable intervals and the algorithm selected one sample from each interval, ensuring that the samples were evenly spaced out across the entire range.
We adopted a logarithmic scaling for the five components in the node $\boldsymbol\kappa$.

We verified that the progenitors were on merging trajectories before evolving the ICs generated by the 47 nodes at full resolution. To do so, we generated the ICs at lower dark matter resolution, $m_{\rm DM}^{low} = 10^3 \unit{M_\odot}$, and evolved them. The coarser resolution makes these simulations run at around a tenth of the computational cost, enabling us to evolve each one for $5 \unit{Gyr}$ and check whether or not the progenitors merge.
In addition, having the same ICs simulated using two different resolutions allows us to evaluate the resolution effects, as is detailed in Appendix \ref{sec:app:resolution_DM}.  

\subsection{Time evolution}\label{sec:sub:numerical_code}

We used the public version of the moving-mesh hydrodynamic code \code{Arepo} \citep{springel_e_2010,weinberger_arepo_2020} to evolve the ICs. 
\code{Arepo} performs the time evolution with a second-order accurate leapfrog scheme, using a tree-particle-mesh method \citep{bagla_treepm_2002} to compute the gravitational interaction between particles.
Having just two galaxies in the simulation box, we adopted only the oct-tree grouping scheme \citep{barnes_hierarchical_1986} for the force evaluation, using nonperiodic boundary conditions.
The gravitational softening length of dark matter and stellar particles is $\epsilon = 1\unit{pc}$, and the simulation box has side $L=100 \unit{kpc}$.

Each simulation evolved until the two progenitors merged into a single, virialized galaxy. The merging instant was identified using the on-the-fly friends-of-friends group finder (FoF) available within \code{Arepo} and originally described in \cite{springel_gadget_2001}. The algorithm groups together particles within a distance of $0.2$ times the mean interparticle separation. We defined the merging instant as the point at which the FoF algorithm finds a single galaxy. 
\commenta{After the merging instant, each simulation evolves for an additional $\sim 1 \unit{Gyr}$, to ensure that the two progenitors fully merge and that the virial ratio is $2T/U \simeq 1$ for several hundred million years, with $T$ being the kinetic energy and $U$ the potential energy.}

\subsection{Merger emulator}\label{sec:sub:merger_emulator}

Having a high computational cost is a common limitation in numerical simulations, especially when a parameter space exploration is involved; a possible solution is the implementation of an emulator, to either reduce the number of simulations needed \citep[e.g.,][]{Bird:2019, Rogers:2019, Brown:2024} or to speed up the computation \citep[e.g.,][]{Spurio:2022, branca:2024, robinson:2024, Bartlett:2024}.
In this work, we have adopted a neural network (NN) as an emulator to enhance the parameter space exploration. The NN approximates a function that, taking the merger configuration and progenitor ICs as input, returns the properties of the post-merger galaxy without requiring new simulations.
Indeed, thanks to the universal approximation theorem \citep[UAT,][]{UAT:1989}, a trained NN can approximate with arbitrary precision the nonlinear function that maps the input parameters, $\boldsymbol\kappa$, into outputs characterizing the galaxy and the stellar halo. For the present work, we set two outputs,
\begin{equation}\label{eq:NN}
   (R_\star, f_5) = {\rm NN}(\boldsymbol\kappa)\,, 
\end{equation}
where with $R_\star$ we refer to the half-mass radius of the post-merger galaxy to distinguish it from the massive progenitor's half-mass radius, $R_{1/2}$, which is one of the inputs of the NN. $f_5$ is the stellar halo mass fraction; that is, the fraction of stars at $r > 5 R_\star$.
Such a choice of outputs enables us to directly compare our results with Tucana~II observations (see Sects. \ref{sec:sub:emulator_results} and \ref{sec:comparison_with_tucana}).

We adopted a regularized loss function to prevent overfitting and trained the emulator on the simulation outputs. The number of training points is small compared to the dimensionality of the problem and we overcame the sparse sampling issue by adopting a Latin hypercube sampling (see Sect. \ref{sec:sub:parameter_space_sampling}), which ensures uniform coverage of the parameter space. Furthermore, deep NNs are relatively unaffected by the growing difficulty of analyzing data as the number of dimensions increases; that is, the \quotes{curse of dimensionality}, when approximating compositional functions \citep{Poggio:2018}. We refer to Appendix \ref{sec:app:neural_netwoks} for a complete description of the NN.


\section{Results}\label{sec:results}

First, we present the evolution of a single simulation and its analysis to overview the merging process (Sect. \ref{sec:merger_overview}).
Then, we repeat the same analysis on all the simulations, use the results to train the emulator, and explore the parameter space to determine the dependency of the stellar halo on the ICs (Sect. \ref{sec:sub:emulator_results}).
Finally, we directly compare our findings with the observations of Tucana II (Sect. \ref{sec:comparison_with_tucana}).

\subsection{Merging of ultra-faint dwarfs: An overview}\label{sec:merger_overview}

\begin{figure}
\centering
\includegraphics[width=0.49\textwidth]{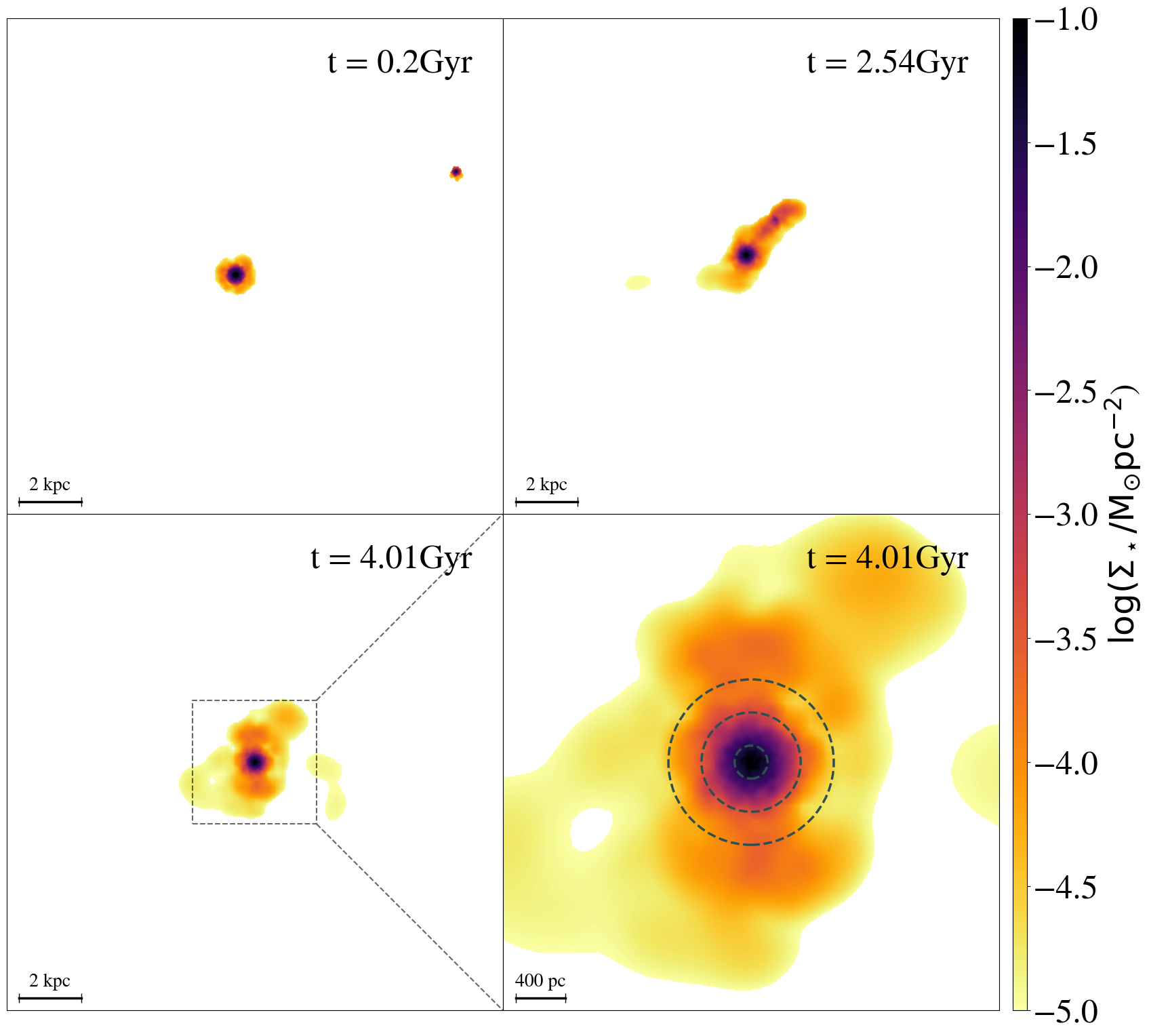}
  \caption{
  Example of merger evolution.
  The different panels report the stellar surface density ($\Sigma_\star$) maps at various evolutionary stages for progenitors with a $M_1/M_2 \simeq 10$ (profile in Fig. \ref{fig:progenitor_IC}) and the merger setup in Fig. \ref{fig:merger_IC}.
  The bottom right panel is a zoom-in on the virialized post-merger galaxy, and the three dashed circles have radii of one, three, and five half-mass radii.
  After the merging, the resulting fraction of stars in the outskirts is $f_5\sim 5.5\%$. 
  \label{fig:merger_evolution_overview}
  }
\end{figure}

\begin{figure}
\centering
\includegraphics[width=0.49\textwidth]{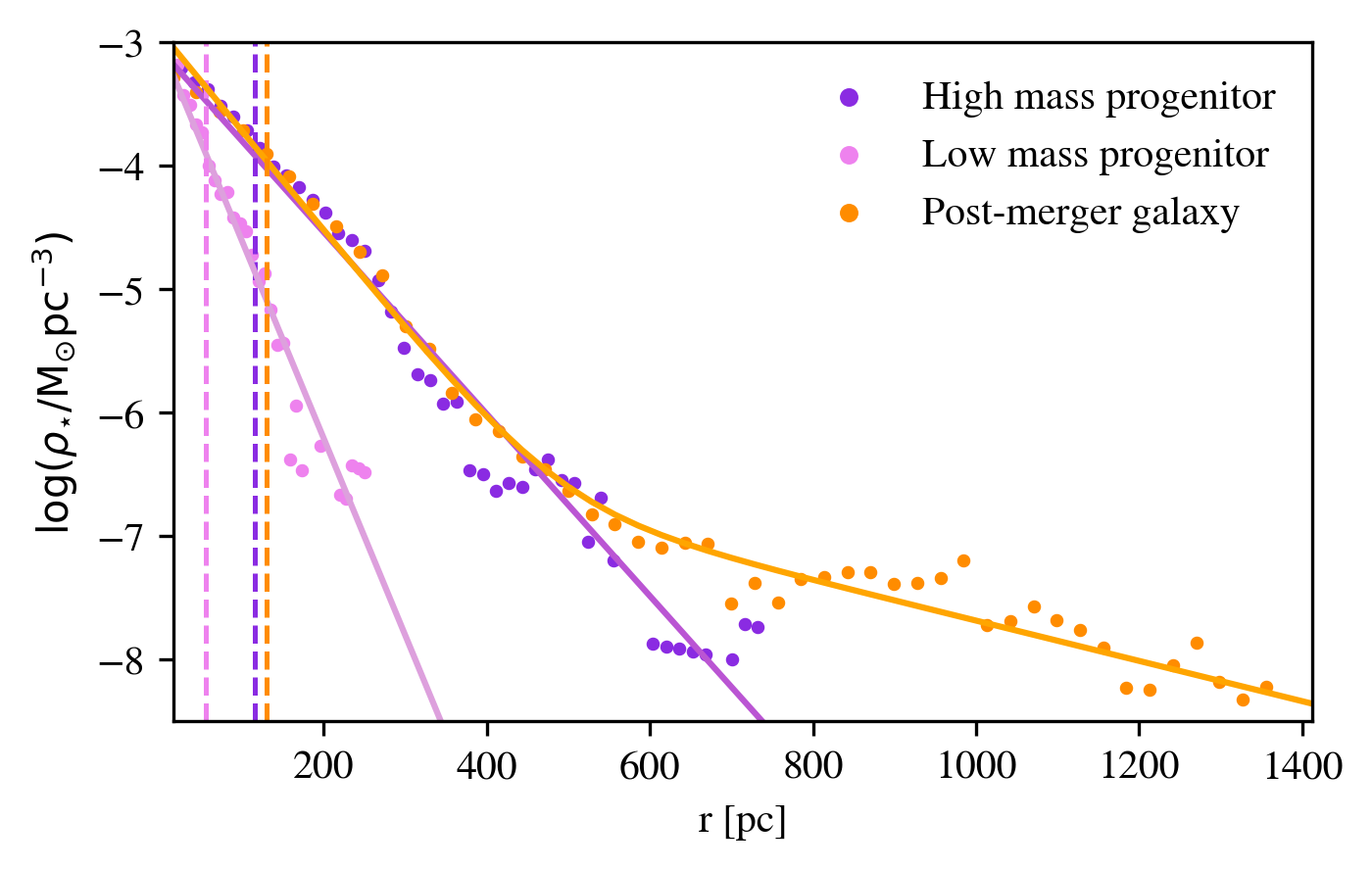}
  \caption{
  Example of pre- vs. post-merging stellar distributions.
  We show the profile for the simulation selected for Fig. \ref{fig:merger_evolution_overview} by reporting $\rho_\star$ for the two progenitors with shades of purple and the post-merger galaxy in orange.
  The solid lines are the fit profile, which is exponential (Eq. \ref{eq:exponentialProfile}) and a sum of exponential for the progenitors and the post-merger galaxy, respectively.
  The vertical lines mark the half-mass radius for each galaxy. 
  \label{fig:density_profile_evolution}
  }
\end{figure}

\begin{figure}
\centering
\includegraphics[width=0.49\textwidth]{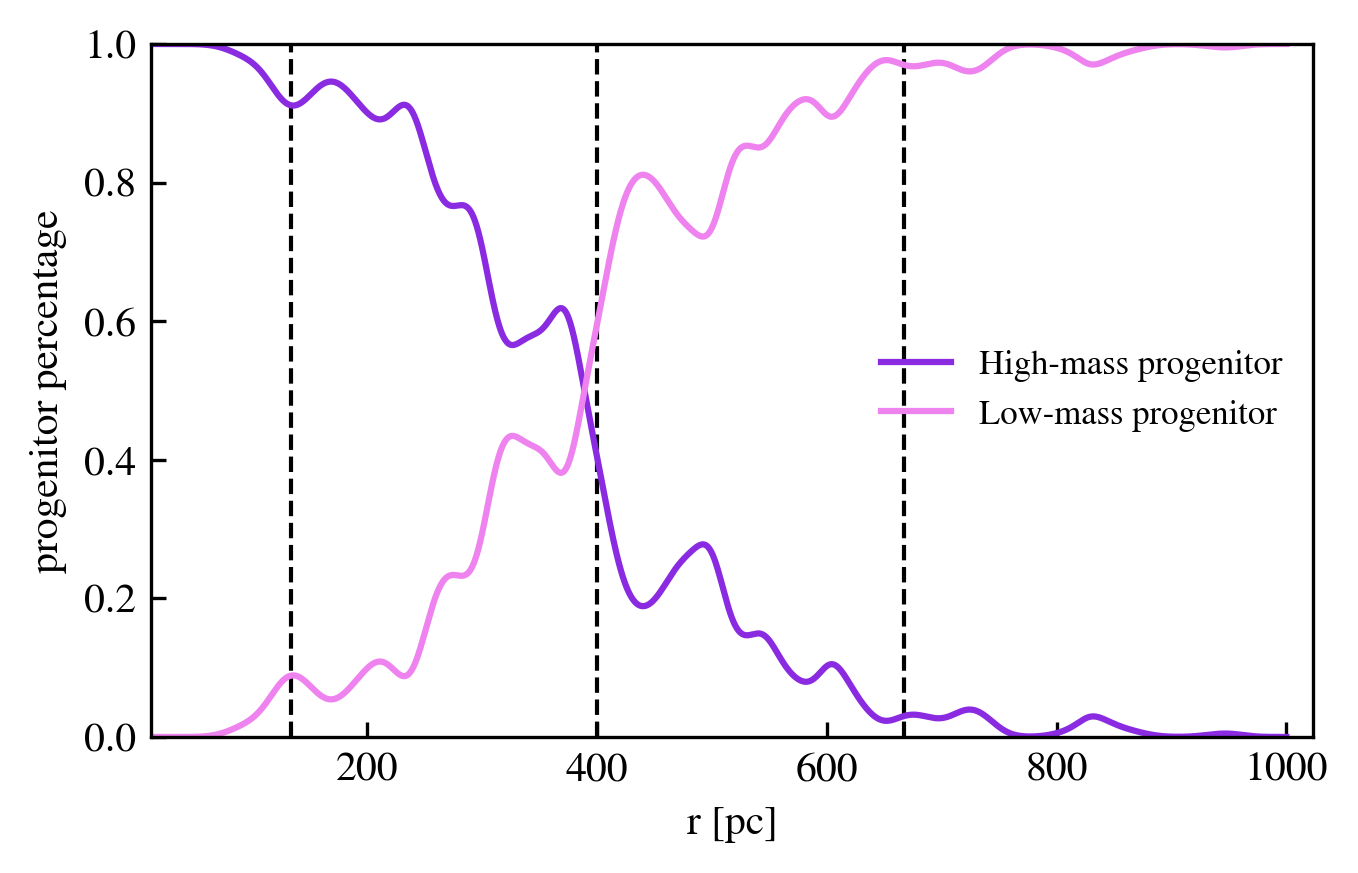}
  \caption{
  Origin of the stars in the post-merger galaxy.
  For the simulation selected for Fig. \ref{fig:density_profile_evolution}, we show the radial distribution of the fraction of stars that originally belonged to the high-mass progenitor (purple) and the low-mass progenitor (pink).
  The dashed vertical lines are at 1, 3, and 5 $R_\star$. 
  \label{fig:density_profile_fraction}
  }
\end{figure}

To showcase the merging process, we selected simulation SIM\_47, which produces a post-merger galaxy with a stellar surface density consistent with the one observed in Tucana~II (see Sect. \ref{sec:comparison_with_tucana}). Simulation SIM\_47 has a merger mass ratio of $M_1/M_2 \simeq 10$, scale radius of $R_{1/2} \simeq 120 \unit{pc}$, mass-to-light ratio of $M_{\rm DM}/M_\star \simeq 10^4 $, specific angular momentum of $l \simeq 1.1 \times 10^3 \unit{pc}^2 \unit{Myr}^{-1}$, and specific kinetic energy of $k \simeq 0.72 \unit{pc}^2 \unit{Myr}^{-2}$. We note that this is the set of parameters already used in Figs. \ref{fig:progenitor_IC} and \ref{fig:merger_IC}, which show the ICs and merger setup, respectively.

To visualize the spatial evolution in the simulated volume, we report the stellar surface density for the various phases of the merger in Fig. \ref{fig:merger_evolution_overview}.
The top left panel is at $t = 0.2 \unit{Gyr}$ when the two progenitors are virialized. The top right panel shows the gravitational interaction between the two progenitors during the third and last passage before merger completion. \commenta{The bottom panels are at the end of the simulation; that is, at $t = 4.01 \unit{Gyr}$, when the post-merger galaxy is virialized, and when the virial ratio is around 1.}
The effect of the merger on the morphology of the galaxy is evident when comparing the post-merger galaxy with one of its progenitors (the right panel in Fig. \ref{fig:progenitor_IC}). To help the comparison, we mark 1, 3, and 5 $R_\star$  in both panels of Fig. \ref{fig:progenitor_IC} and Fig. \ref{fig:merger_evolution_overview}. The progenitor is mostly contained within 5 $R_{1/2}$, while the post-merger galaxy is wider, extending way beyond 5 $R_\star$. The extended stellar distribution at large radii -- the stellar halo -- forms as a result of the merger event and is bar-shaped.
Therefore, contrary to the progenitors, the post-merger galaxy is asymmetric and we account for the asymmetry using an ensemble average\footnote{We consider all possible projection planes, take the median as $R_{1/2}$ and the 16-th and 84-th percentiles as the error.} to estimate $R_\star$, which we recall is the half-mass radius of the 2D projection of the galaxy. \commenta{We address the differences between the 3D and 2D half-mass radius in Appendix \ref{sec:app:projections}.}

To analyze the difference between pre- and post-merger galaxies' stellar distribution, in Fig. \ref{fig:density_profile_evolution} we compare the stellar density of the two progenitors (shades of purple) and the post-merger galaxy (orange).
The vertical lines are the half-mass radius for each galaxy, while the solid lines are the fit profiles: exponential (Eq. \ref{eq:exponentialProfile}) for both progenitors and the sum of two exponential profiles for the post-merger galaxy. The double exponential profile models the formation of a dense core and a sparse stellar halo and better fits the post-merger stellar distribution.
The difference between the post-merger galaxy and its massive progenitor stellar distribution lies in the outskirts, where the post-merger galaxy has a higher mass fraction due to the formation of a stellar halo. The higher mass fraction affects the half-mass radius, going from $R_{1/2} =( 114\pm 1) \unit{pc}$ of the massive progenitor to $R_\star = 130^{+3}_{-1} \unit{pc}$ of the post-merger galaxy. The error distribution in $R_\star$ is due to the asymmetry of the stellar halo formed via a merger.  

Tidal interactions during the merger strip stars from the less massive progenitor and form the stellar halo, as is shown in Fig. \ref{fig:density_profile_fraction}, where we report the fraction of stars of the post-merger galaxy coming from each progenitor at different radii. We adopted the same progenitor colors of Fig. \ref{fig:density_profile_evolution}, and the dashed lines are at 1, 3, and 5 $R_\star$.
While the center of the post-merger galaxy is dominated by stars of the more massive progenitor, beyond 3 $R_\star$ the majority of the stars come from the smaller progenitor. This distribution shows that stars of the low-mass progenitor compose the stellar halo, and this is key to understanding the trends we observe exploring the parameter space (Sect. \ref{sec:sub:emulator_results}).      

To quantify the presence of a stellar halo, we followed \citet{Tarumi:2021} and calculated $f_5$, which we recall is the fraction of stars beyond 5 $R_{1/2}$. Since $f_5$ depends on the half-mass radius, we performed the same ensemble average considering all possible projection planes. For each projection plane, we randomly sampled the stellar surface density distribution using the same number of stars of \citet[][i.e. 19]{chiti:2021} and considered the mean and standard deviation as $f_5$ and its error. \commenta{We refer the reader to Appendix \ref{sec:app:projections} for a discussion on the differences between 3D and projected properties.}
We estimate $f_5 = 0.16 \pm 0.4\%$ and $f_5 = 0.19\pm 0.03\% $ for the two progenitors. These values are consistent with the ones expected in isolated galaxies with an exponential profile, where $f_5$ tends to zero. 
Figure \ref{fig:merger_evolution_overview} shows the formation of a stellar halo after the merger, and as a result of the stellar halo formation, the fraction of outskirts stars in the post-merger galaxy increases to $f_5 = 5.4\pm 1.8\% $.
For comparison, the observed value in Tucana~II is $f_5^{\rm TucII} = 2/19 = 10.5 \%$. 

\subsection{Exploration of the parameter space with the emulator}\label{sec:sub:emulator_results}

\begin{figure*}
\centering
\includegraphics[width=0.99\textwidth]{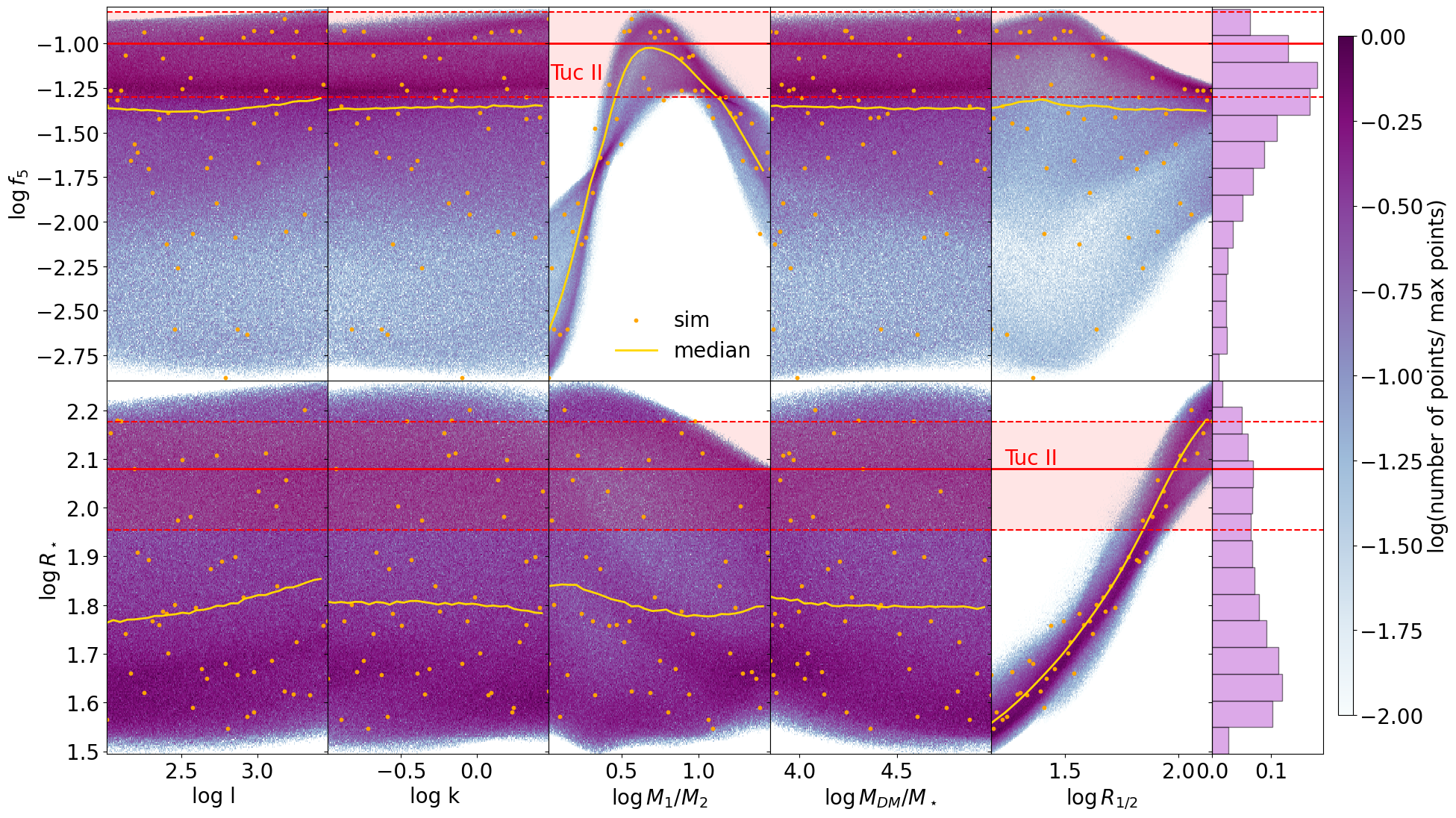}
  \caption{Emulation of the post-merging $f_5$ (upper panels) and stellar radius ($R_\star$, lower panels) as a function of merger setup parameters, i.e., specific angular momentum ($l$, Eq. \ref{eq:def:specific_L}), specific kinetic energy ($k$, Eq. \ref{eq:def:specific_K}), mass ratio ($M_1/M_2$), dark to stellar mass ratio ($M_{\rm DM}/M_\star$), and pre-merger stellar size ($R_{1/2}$, \ref{eq:exponentialProfile}).
  Each of the first five columns gives the distribution of outputs of the merging process as a function of a single parameter. The distribution is obtained by emulating $10^6$ IC log-spaced distributed (see Tab. \ref{table:parameterSpace_intervals}), and the gold line represents its median.   
  The orange points are the individual simulations used to train the emulator.
  The red line (shaded region) gives the average (dispersion) of the observed values for Tucana~II \citep{Bechtol:2015,chiti:2021}. 
  The last column shows the distribution of the emulated output normalized to the total number of inferences. 
  \label{fig:merger_input_vs_output}
  }
\end{figure*}

Using the radius, $R_\star$, and the fraction of stars in the outskirts, $f_5$, of each post-merger galaxy, we built the dataset to train the emulator. We recall that (Sect. \ref{sec:sub:merger_emulator}) the input values of the emulator are the explored initial parameters, $l$, $k$, $M_1/M_2$, $M_{\rm DM}/M_\star$, and $R_{1/2}$, and the output values are the median of the two post-merger quantities, $f_5$ and $R_\star$ (see Eq. \ref{eq:NN}; for details of the NN architecture and its training, we refer to Appendix \ref{sec:app:neural_netwoks}).

Figure \ref{fig:merger_input_vs_output} shows the distribution of the emulated outputs ($f_5$ top row, $R_\star$ bottom row) of $10^6$ different ICs log-spaced distributed. The orange dots represent the results of the training simulations, and their overlap with the emulated results confirms the emulator's robustness. 
The last column of Fig. \ref{fig:merger_input_vs_output} shows the distribution of output values in the parameter space explored. The distribution of $f_5$ peaks around $5\%$, the value found by \citet{Tarumi:2021} for their dry merger. The $R_\star$ distribution increases at low values. This trend is primarily due to the distribution of the training data, which is affected by numerical heating at low $R_{1/2}$, as is explained in detail in Appendix \ref{sec:app:resolution_DM}. Consequently, the emulator predicts a higher frequency of galaxies with low $R_\star$ due to the higher number of simulations with $R_\star \simeq 40 \unit{pc}$. 

The red-shaded areas are the observed values for Tucana~II with the relative error. 
\commenta{There are several estimates of the angular size of Tucana~II \cite[][]{Bechtol:2015, Koposov:2015, Moskowitz:2020} and distances \citep{Bechtol:2015, Koposov:2015, Vivas:2020} each resulting in different half-light radii. To ensure consistency with \citet{chiti:2021}, we adopted the half-light radius $R^{\rm TucII}_{1/2} = (120 \pm 30) \unit{pc}$ \citep{Bechtol:2015}. This choice allows for a direct comparison of our simulated stellar distribution with their observational data.}
We computed the fraction of distant stars, $f_5^{\rm TucII} = 10 \pm 5 \%$, from the distribution of individual stars observed in \cite{chiti:2021}. The error assigned to $f_5$ considers that one of the two stars in the outskirts can lie within 5 $R_{1/2}$, changing the fraction to $5\%$. 

The emulated results, whose median is given by the gold line, show little to no dependency on the specific angular momentum, $l$, kinetic energy, $k$, and dark matter content, $M_{\rm DM}/M_\star$.
Clear trends are instead present for the merger mass ratio, $M_1/M_2$, and the progenitor half-mass radius, $R_{1/2}$; specifically, $R_\star$ correlates almost linearly with $R_{1/2}$, while $f_5$ increases with $M_1/M_2$, peaks around $M_1/M_2 \simeq 6$, and then decreases. 
The correlation between the sizes of the progenitor and the post-merger galaxy is expected, since the larger the progenitor is, the larger the post-merger galaxy will be.
Meanwhile, the correlation between $f_5$ and the merger mass ratio emerges because, as is discussed in Sect. \ref{sec:merger_overview}, the stellar halo stars primarily originate from the less massive progenitor (the \quotes{satellite}), leading to two distinct regimes. When the two progenitors have comparable masses, they mix, forming a post-merger galaxy with a minimal stellar halo and a small $f_5$. In this regime, increasing the merger mass ratio weakens the depth of the potential well of the satellite, making it easier for its stars to be stripped during the merger and form a more prominent stellar halo. As a result, $f_5$ initially increases with the $M_1/M_2$ ratio. However, increasing $M_1/M_2$ reduces the number of stars in the \quotes{satellite}, leading to a decrease in the number of stars available to form the stellar halo. For this reason, beyond a certain threshold (approximately $M_1/M_2 \simeq 6$), $f_5$ begins to decline.     

To quantify the dependencies of $f_5$ and $R_\star$ from the ICs ($\boldsymbol{\kappa}$), we performed a principal component analysis (PCA). A PCA studies the linear transformation (component) of the parameter space of the data. Such components are ranked so that the variance decreases. Thus, the first (principal) components can explain most of the variation in the original dataset.
We performed the PCA separately for the two outputs; that is, one for $(f_5, \boldsymbol{\kappa})$ and one for $(R_\star, \boldsymbol{\kappa})$.
For each PCA, we considered $10^6$ points in six dimensions, normalizing each dimension such that its mean and standard deviation are 0 and 1, respectively.
The first PCA component for $f_5$ ($R_\star$) is dominated by $f_5$ ($R_\star$) and $M_1/M_2$ ($R_{1/2}$), confirming the trends observed in Fig. \ref{fig:merger_input_vs_output}.

To determine the secondary dependencies, we performed two additional PCAs considering a reduced number of points. Specifically, when analyzing $f_5$, we considered the points with $M_1/M_2 \in [5,7]$, where $f_5$ peaks, while for $R_\star$, we considered the points with the progenitor radius $R_{1/2} \in [90,110] \unit{pc}$; that is, the progenitor radius producing a post-merger galaxy radius consistent with Tucana~II observations.
The first reduced PCA component for $f_5$ ($R_\star$) is dominated by $f_5$ ($R_\star$) and $R_{1/2}$ ($M_1/M_2$), revealing that the principal dependency of one output is the secondary dependency of the other. 

Once we removed the dependency on $M_1/M_2$ and $R_{1/2}$, selecting points simultaneously in the intervals $M_1/M_2 \in [5,7]$ and $R_{1/2} \in [125,145] \unit{pc}$, we noticed that $f_5$ and $R_\star$ marginally depend on the other three parameters, $l$, $k$, and $M_{\rm DM}/M_\star$. The relative importance depends on the intervals considered, but generally $M_{\rm DM}/M_\star$ is the least important for both $f_5$ and $R_\star$.

\subsection{Comparison with Tucana II observations}\label{sec:comparison_with_tucana}

\begin{figure}
\centering
\includegraphics[width=0.45\textwidth]{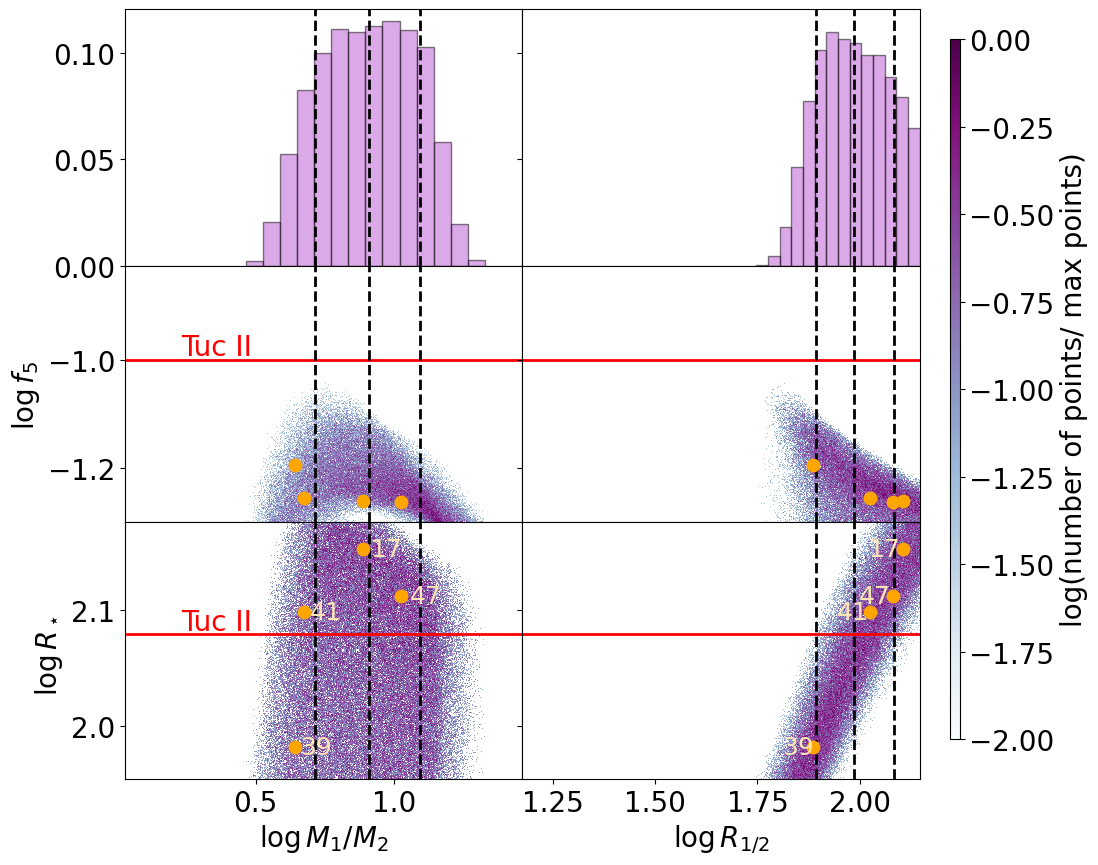}
  \caption{Distributions of the input couple $R_{1/2}$ and $M_1/M_2$ that give $f_5$ and $R_\star$ simultaneously consistent with the observations from Tucana~II.
  Similarly to Fig. \ref{fig:merger_input_vs_output}, the red line gives the average value inferred from observations, but the ranges for $f_5$ and $R_\star$ were cut to include only the observational uncertainty, i.e., $f_5^{\rm TucII} = 10 \pm 5 \%$ \citep{chiti:2021} and $R^{\rm TucII}_{1/2} = (120 \pm 30) \unit{pc}$ \citep{Bechtol:2015}, respectively. 
  The orange points are the post-merger galaxies consistent with observations from Tucana~II, and are numbered with their simulation ID.
  \label{fig:zoom_merger_input_vs_output}
  }
\end{figure}

\begin{figure*}
\centering
\includegraphics[width=0.99\textwidth]{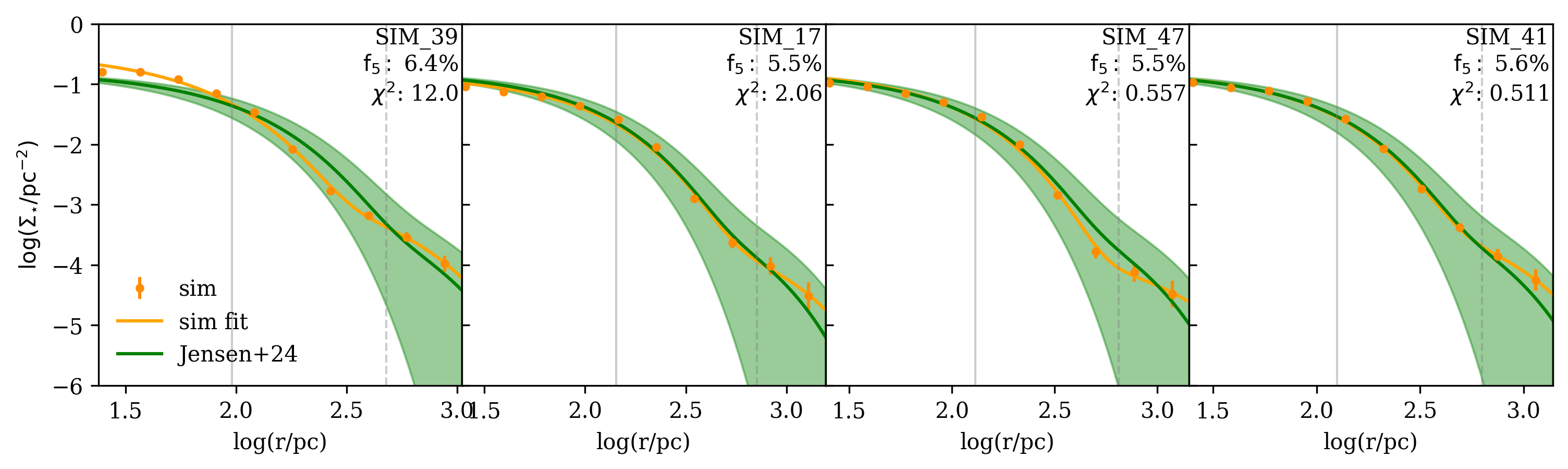}
  \caption{Comparison of the observed stellar number density distribution of Tucana~II and the simulations with the $f_5$ and $R_{1/2}$ parameters closest to the observed values ($f_5^{\rm TucII} = 10 \pm 5 \%$ from \citealt{chiti:2021} and $R^{\rm TucII}_{1/2}/\unit{pc} = 120 \pm 30$ from \citealt{Bechtol:2015}).
  Each panel shows a different simulation, as is indicated in the inset, ordered by decreasing $\chi^2$ value.
  The green line (shaded area) gives the mode (error) of the profile for Tucana~II, observed by \citet{Jensen:2024}.
  The orange line (points with bars) gives the best fit (average and standard deviation of the data points) of the profile of a simulated post-merger galaxy, obtained by adopting a two-exponential profile (Eq. \ref{eq:DoubleExponentialProfile}).
  The solid (dashed) gray line shows the $R_\star$ (5 $R_\star$) post-merger radius.     
  \label{fig:jensen_comparison}
  }
\end{figure*}

We can now compare the emulated $f_5$ and $R_\star$ with the observations to infer the properties of Tucana~II progenitors.
As is described in Sect. \ref{sec:sub:emulator_results}, we adopted $R^{\rm TucII}_{1/2} = (120 \pm 30) \unit{pc}$ and $f_5^{\rm TucII} = 10 \pm 5 \%$ as observed values of half-mass radius and fraction of stellar halo stars in Tucana~II, respectively.

Figure \ref{fig:zoom_merger_input_vs_output} shows the distribution of the output values consistent with Tucana~II observations, and based on the PCA results, we limited the study to two initial parameters that define the output values; that is, $M_1/M_2$ and $R_{1/2}$. The top row shows the distribution of these initial parameters, with the dashed lines indicating the median and 16-th and 84-th percentiles, while the four orange points represent four simulations in the suite consistent with Tucana II observations.
Consistency with observations is achieved for ICs with a progenitor half-mass radius of $R_{1/2} = 97^{+25}_{-18} \unit{pc}$ and a merger mass ratio of $M_1/M_2 = 8 _{-3}^{+4}$. Therefore, we determine that an intermediate merger ($5<M_1/M_2<12$) forms a stellar halo similar to the one observed for Tucana~II. Such a finding is consistent with \citet{Deason:2014}, who report that 40\% of $z=0$ UFDs in their simulations underwent at least one major merger in the last $12 \unit{Gyr}$. 
Among the four post-merger simulations consistent with observations, SIM\_41 and SIM\_47 are consistent within $1.1\sigma$ with\footnote{For the results of our models, the left (right) $\sigma$ is defined as the interval between the 16-th (84-th) percentile and the median. Different definitions of $\sigma$ give slight quantitative differences, but the physical picture is unchanged.} the inferred values of $M_1/M_2$ and $R_{1/2}$, SIM\_17 has a slightly larger progenitor radius, $>1.1\sigma$, and SIM\_39 has both $M_1/M_2$ and $R_{1/2}$ values $>1.1\sigma$. 
We note that the errors on the inferred properties highly depend on the value and error adopted for $f_5$, and better constraining it helps to decrease the uncertainties on $M_1/M_2$ and $R_{1/2}$. For example, lowering $f_5$ error by a factor of two -- $\sim \pm 2.5 \%$ -- results in a decrease of a factor of three on the error of the ICs consistent with Tucana~II. 

One method to reduce the error on $f_5$ is to increase the number of identified stars; for example, using Gaia data. In recent work, \citet{Jensen:2024} exploit Gaia data, employing a selection algorithm based on spatial, color-magnitude, and proper motion information targeting low brightness features of MW dwarf satellites. They are able to identify the second low-density outer profile of nine satellites, including Tucana~II, and for each of them, to derive the stellar surface density profile. They model it as the sum of two exponential profiles:
\begin{equation}\label{eq:DoubleExponentialProfile}
    \Sigma(r) = \Sigma_{0} \left( e^{-r/ r_e} + B e^{-r/r_s}\right)\,,
\end{equation}
where $\Sigma_{0}$ is the central surface density, $r_e$ is the exponential scale radius, $B$ is the normalization of the outer component ($0\leq B \leq 1$), and $r_s$ the scale radius of the outer component ($1.68 r_s > R_{1/2}$). They derive $r_e$ from the literature value of the half-light radius such that $R_{1/2} = 1.68r_e$. 
Unfortunately, we cannot use their star candidates list to evaluate $f_5$, since there are no spectroscopic follow-ups to verify the membership of each star. However, we can test our results by comparing the surface density of our simulated post-merger galaxies with the one derived by them. 

Figure \ref{fig:jensen_comparison} shows the derived stellar surface density profile of Tucana~II (green) and our simulated post-merger galaxies (orange), and each of the four columns refers to a simulation that is simultaneously consistent with the observed values of $f_5$ and $R_{1/2}$. 
As we can see, most of these four simulations reasonably match the observed stellar halo well beyond $1 \unit{kpc}$ and only SIM\_39 is not consistent with the observed data. We quantified the accordance between simulations and observations using the standard $\chi^2$ procedure:
\begin{equation}\label{eq:chi-square}
    \chi^2 = \sum_{i=1}^{10} \frac{\left(\Sigma_{i,obs} - \Sigma_{i,sim}\right)^2}{ \sigma_{i,obs}^2+\sigma_{i,sim}^2}\,,
\end{equation}
where $\Sigma_{i,obs}$ ($\sigma_{i,obs}$) and $\Sigma_{i,sim}$ ($\sigma_{i,sim}$) are the stellar surface density profiles (standard errors) derived from observations and simulations, respectively. We used ten radius bins to be consistent with \citet{Jensen:2024}, and for each bin we computed $\sigma_{i,obs}$ propagating $R_{1/2}$, $B$, and $r_s$ errors, and we computed $\sigma_{i,sim}$ as the Poisson error. 
The three simulations within $1.1\sigma$ from the inferred $M_1/M_2$ -- SIM\_17, SIM\_41, and SIM\_47 -- have a good agreement; that is, $\chi^2 = 0.5- 2$. It should be noted that $\chi^2$ of SIM\_17 is larger, since the accordance with $R_{1/2}$ is worse. All three simulations show the two-component profiles with differences in the central density $\Sigma_{0}$ and the transition to the second exponential profile. Their ICs span almost all the angular momentum values, confirming that the formation of a stellar halo does not depend on angular momentum in the interval explored in this work. We recall that we considered $l$ values that result in a merger between the two progenitors.
For SIM\_39, the surface density significantly diverges from the observed one ($\chi^2 = 12.0$). Although the fraction of distant stars is consistent with observations, $f_5 = 6.4^{+2}_{-4}$, the simulation features an inner core that is $2\times$ denser and $0.5\times$ smaller compared to observations. This is due to the offset from $M_1/M_2$ and $R_{1/2}$, which are at >1.1 $\sigma$ from the emulator predictions.

\section{Summary and discussion}\label{sec:summary_and_discussion}

In this work, we investigate merger events as the formation scenario of stellar halos around UFDs. We study the impact of different merger and progenitor properties and focus on reproducing the stellar halo observed in Tucana~II by \citet{chiti:2021,chiti:2023, Jensen:2024}. 

To this end, we developed a suite of 47 N-body simulations of isolated dry mergers between two idealized progenitors of Tucana~II with a stellar resolution of a single solar mass. We used \code{DICE} \citep{perret_dice_2016} to generate the ICs, assuming an exponential profile for the stellar component of both progenitors, and we used \code{Arepo} \citep{weinberger_arepo_2020} to evolve the simulation until the post-merger galaxy virializes. The suite explores different merger trajectories and progenitors properties, assigning to each simulation a unique set of five parameters: i) the specific angular momentum, $l$, ii) the specific kinetic energy, $k$, iii) the merger mass ratio, $M_1/M_2$, iv) the dark matter content, $M_{\rm DM}/M_\star$, and v) the half-mass radius of the more massive progenitor, $R_{1/2}$. For each post-merger galaxy, we quantified the formation of a stellar halo using two properties: the half-mass radius, $R_\star$, and the fraction of stars, $f_5$, at high radii (>$5 R_\star$).

We developed a NN-based emulator to avoid the extreme computational cost expected from a high-resolution sampling of the selected five-dimensional parameter space. Via the emulator, we quickly obtained the properties of one million post-merger galaxies -- $R_\star$ and $f_5$ -- and studied the dependency of the stellar halo formation on the ICs of the merger and progenitors. By comparing the emulated properties to Tucana II observations, we inferred consistent merger and progenitor characteristics. Additionally, we compared the stellar surface density of simulated post-merger galaxies and Tucana~II to further validate the merger scenario.   
           
We can summarize our most important results as follows:
\begin{itemize}
    \item[$\bullet$] The merger mass ratio, $M_1/M_2$ (progenitors' size, $R_{1/2}$), drives the formation of a stellar halo, $f_5$ (the size of the galaxy, $R_\star$), in the aftermath of a merger, with the progenitors' size, $R_{1/2}$ (merger mass ratio, $M_1/M_2$), being the secondary dependency. In the interval explored, the specific angular momentum, $l$, specific kinetic energy, $k$, and dark matter abundance, $M_{\rm DM}/M_\star$, have a negligible impact on both the stellar halo formation and the size of the post-merger galaxy.
    \item[$\bullet$] Using the observed fraction of distant stars, $f_5^{\rm TucII}= 10.5\%$ \citep{chiti:2021}, and half-light radius, $R_\star^{\rm TucII} = 120 \unit{pc}$ \citep{Bechtol:2015}, of Tucana~II, we predict that the merger mass ratio and massive progenitor size consistent with observations are $M_1/M_2 = 8_{-3}^{+4}$ and $R_{1/2} = 97^{+25}_{-18} \unit{pc}$. The merger that can form Tucana~II's stellar halo is intermediate, while the size of present-day Tucana~II is similar to the radius of its massive progenitor.  
    \item[$\bullet$] Four simulated post-merger galaxies are consistent with $f_5^{\rm TucII}$ and $R_\star^{\rm TucII}$. Three simulations have $M_1/M_2$ ($R_{1/2}$) within $1.1\sigma$ ($1.2\sigma$) from the predicted values: their stellar surface density reproduce ($\chi^2 \approx 0.5 -2$) the observed one \citep{Jensen:2024} well. The simulation with both $M_1/M_2$ and $R_{1/2}$ at $> 1.2\sigma$ from the predictions deviates the most from observations ($\chi^2 = 12$). This underlines the importance of $M_1/M_2$ and $R_{1/2}$ in determining the post-merger galaxy morphology.
\end{itemize}

Our results confirm previous findings about the importance of the merger mass ratio on the formation of a stellar halo in UFDs \citep[][]{deason:2022, Ricotti:2022}, show the impact of the progenitor size on the radius of the post-merger galaxy, and highlight the possibility of inferring progenitor and merger properties using present-day observations.
The agreement between the results from our simulations and emulations with observations in terms of $f_5$, $R_\star$, and stellar surface density further corroborates the merging scenario for the formation of Tucana~II's stellar halo, suggested by observations \citep{chiti:2021, chiti:2023} and simulations of two specific UFD progenitors \citep{Tarumi:2021}. 

\commenta{The NN-based emulator proves to be a fundamental tool to perform the parameter space exploration, especially when computational resources are limited, and it can be easily adapted to infer any scalar property from simulations; for example, the post-merger galaxy's total mass. To emulate functional properties, such as the parametrical dependence of density profiles or surface brightness, and directly compare them with observations \citep[e.g.][]{Jensen:2024, Conroy:2024}, it is required to change the architecture to a neural operator-based model \citep{Lu:2019}. Such a possibility highlights the versatility and power of emulation techniques in astrophysical research.}

We expect ongoing and future spectroscopic surveys targeting UFDs to have a high impact on the search for stellar halos. For example, \code{4dwarfs} \citep{Skuladottir:2023} will provide the chemical abundances for more than 2000 stars in $\sim 40$ UFDs (Sk{\'u}lad{\'o}ttir, private communication). The higher statistic will produce precise measurements of $f_5$ in Tucana~II and possibly other UDFs, revealing the frequency of stellar halos around UFDs. Using our method on this upcoming data will allow us to uncover the past assembly history of the faintest companions of our MW. 


\begin{acknowledgements}
This project received funding from the ERC Starting Grant NEFERTITI H2020/804240 (PI: Salvadori).
We acknowledge the CINECA award under the ISCRA initiative for the availability of high-performance computing resources and support from the Class C project UFD-SHF, \quotes{UFD mergers: Stellar Halo Formation}, HP10CM106H (PI: Querci).
We acknowledge the usage of
     \code{Arepo} \citep{weinberger_arepo_2020},
     \code{astropy} \citep{The_Astropy_Collaboration_2022},
     \code{DeepXDE} \citep{lu2021deepxde},
     \code{DICE} \citep{perret_dice_2016},
     \code{matplotlib} \citep{hunter_matplotlib_2007},
     \code{numpy} \citep{harris_array_2020},
     \code{pynbody} \citep{ponzen_pynbody_2013},
     \code{python3} \citep{Python3},
     \code{scipy} \citep{Scipy:2020},
     and \code{scikit learn} \citep{scikit-learn}.
\end{acknowledgements}

\bibliographystyle{aa_mod}
\bibliography{master}{}
    
\begin{appendix}

\FloatBarrier
\section{Dynamical heating from dark matter resolution} \label{sec:app:resolution_DM}

\commenta{Two species simulations, i.e. dark matter and stars, are affected by numerical heating that depends on the resolution mass ratio $\mu = m_{\rm DM}/m_{\star}$ and on the number of particles \citep{Ludlow:2023}. High $\mu$ or low number of particles induce numerical heating, resulting in spurious heating of the stellar component and a larger galaxy.}
\commenta{To test the impact of numerical heating on our results, we compare the properties of the post-merger galaxies of three suites, varying the dark matter and stellar resolution. The dark matter particle of the low-resolution (high-resolution) suite has a mass of $m_{\rm DM}^{low} = 10^3 \unit{M_\odot}$ ($m_{\rm DM}^{high} = 10^2 \unit{M_\odot}$) and the resolution mass ratio is $\mu^{low} =  10^3 \unit{M_\odot}$ ($\mu^{high} = 10^2$). The mass of a single dark matter particle in the mid-resolution suite is $m_{\rm DM}^{mid} = m_{\rm DM}^{low}$ while $\mu^{mid} = \mu^{high}$. The adopted $\mu^{low}$ is similar to the value used by \citet{agertz:2020}, while $\mu^{high}$ is consistent with the value in \citet{Ricotti:2022}.}

\begin{figure}[h!]
\centering
\includegraphics[width=0.45\textwidth]{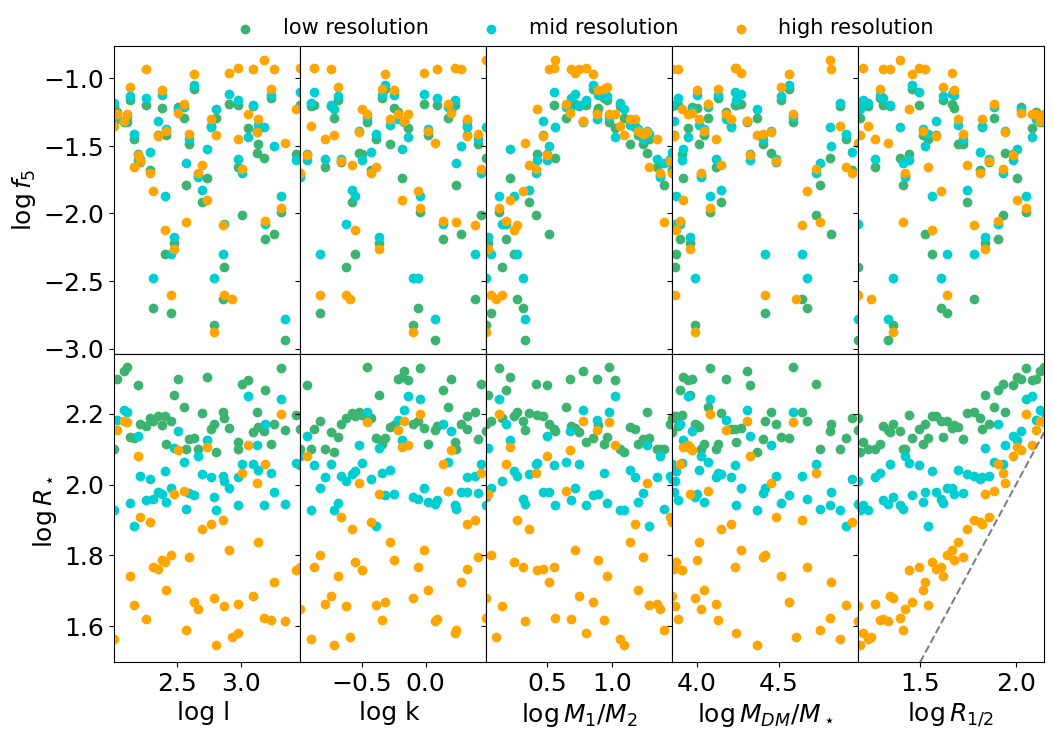}
  \caption{Comparison of the distribution of the post-merger galaxies properties, i.e., half-mass radius $R_\star$ (bottom row) and the fraction of stars $f_5$ beyond $5 R_\star$ (top row), changing the dark matter and stellar resolution. 
  Each column represents an IC property explored in this work, from left to right are: specific angular momentum $l$, specific kinetic energy $k$, merger mass ratio $M_1/M_2$, dark matter abundance $M_{\rm DM}/M_\star$, and progenitor half-mass radius $R_{1/2}$. 
  The dashed line in the bottom right corner is $R_\star = R_{1/2}$.
  \commenta{The low-resolution simulations ($m_{\rm DM} = 10^3 \unit{M_\odot}$, $m_{\rm DM}/m_\star = 10^3 \unit{M_\odot}$) are in green, the mid-resolution simulations are in blue ($m_{\rm DM} = 10^3 \unit{M_\odot}$, $m_{\rm DM}/m_\star = 10^2 \unit{M_\odot}$), while the high-resolution simulations ($m_{\rm DM}^{high} = 10^2 \unit{M_\odot}$, $m_{\rm DM}/m_\star = 10^2 \unit{M_\odot}$) are in orange.}
  \label{fig:resolution_comparison}
  }
\end{figure}

In Fig. \ref{fig:resolution_comparison}, we show the distribution of the post-merger galaxies properties, $f_5$ top row and $R_\star$ bottom row, in the parameter space explored. The high-resolution simulations are in orange\commenta{, the mid-resolution in blue, }and the low-resolution in green. 
Comparing the distributions of the fraction of distant stars, we note that they overlap, with the higher resolution suite peaking at a higher value ($\sim13 \%$) compared to \commenta{the mid-resolution ($\sim 8.6\%$) and} the low-resolution ($\sim 8.2 \%$). 
On the contrary, the half-mass distributions do not overlap, with the \commenta{mid-resolution $R_\star$ systematically between the low- and high-resolution one.} 

Considering the typical error estimating $f_5$, i.e., of the order of 20\%, the $f_5$ distributions at different resolutions are consistent. Such consistency confirms that the simulations capture the stellar halo formation by a dry merger, even with a resolution as low as $m_{\rm DM}^{low}$. On the contrary, the half-mass radius distributions do not overlap, even considering the error associated with the measure, showing the impact of spurious numerical heating and forcing a different method to test the convergence of the simulations.
The size of the post-merger galaxy (orange points) is larger than the progenitor radius (the dashed line), as expected in N-body dry-merger simulations \citep{Oogi:2016} and evident from the bottom right panel of Fig. \ref{fig:resolution_comparison}. For $R_{1/2} > 50 \unit{pc}$, the two radii are proportional, while below 50 pc $R_\star$ saturates at a minimum value, i.e., the radius resolution limit, suggesting that the numerical heating can be neglected in simulations with $R_{1/2} > 50 \unit{pc}$.  
To further verify that numerical heating is negligible, we limit $R_{1/2}$ to the proportional regime and determine the impact of each parameter on $R_\star$. We find the same hierarchy as when using the whole dataset, with $R_{1/2}$ being the dominant parameter and $M_1/M_2$ playing a secondary role, showing that numerical heating does not affect our results. Moreover, the values of $R_\star$ derived at $R_{1/2} \simeq 100 \unit{pc}$, i.e., the expected value of the massive progenitor radius consistent with observations, for different resolutions are consistent when considering the observational errors are of the order of $\sim 30 \unit{pc}$. 

For the reasons stated above, our results on the stellar halo formation and size of the post-merger galaxies are well captured by the dark matter resolution $m_{\rm DM}^{high} = 10^2 \unit{M_\odot}$, and there is no need for increasing the resolution.           

\FloatBarrier
\section{Using neural networks as an emulator} \label{sec:app:neural_netwoks}

In this work, we adopt and train a NN as an emulator to infer an approximate value of $f_5$ and $R_\star$ of post-merger galaxies based on the ICs of the merger.
The NN has an input layer with 5 neurons, one for each parameter explored, i.e., $l$, $k$, $M_1/M_2$, $M_{\rm DM}/M_\star$, and $R_{1/2}$, 5 hidden layers with 128 dense neurons, and an output layer with a neuron for $f_5$ and $R_\star$, each.

The number of simulations available for training and testing the NN is limited, and each simulation carries information in a specific interval of ICs. To avoid losing information, we train the NN on all the available data, and we use a l2 regularization to control overfitting \citep{buhlmann2011statistics}.
We tested our approach by training nine identical emulators on 90\%, 80\%, and 70\% of the dataset and testing them on the remaining fraction.

\begin{figure}[h!]
\centering
\includegraphics[width=0.45\textwidth]{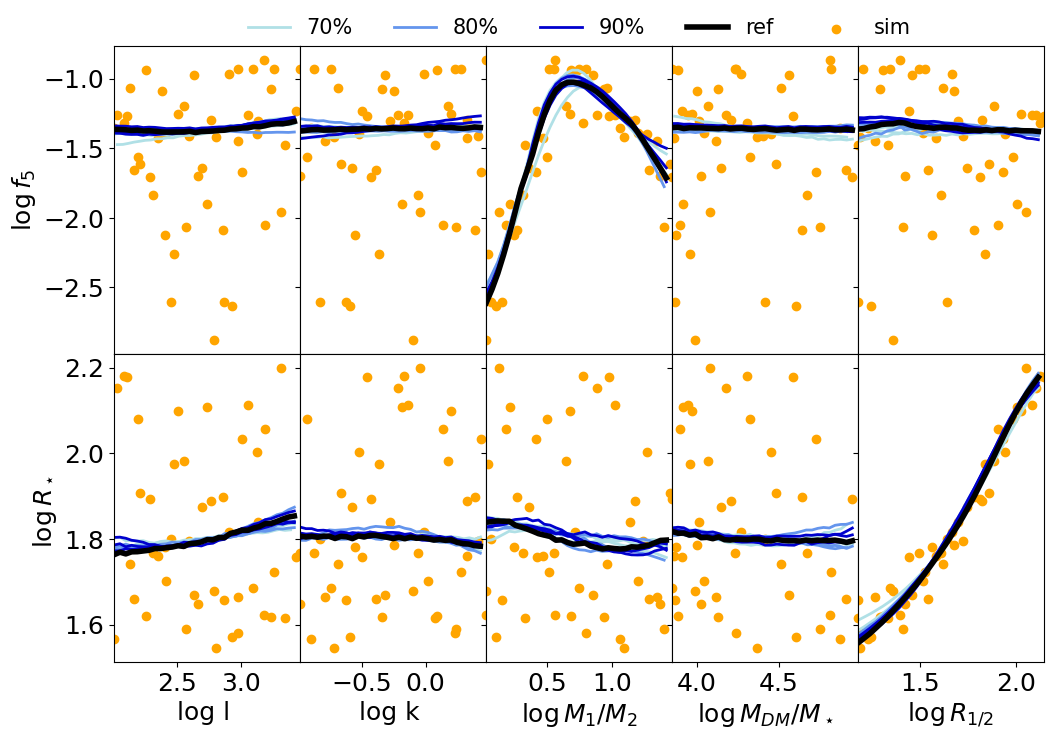}
  \caption{Same as Fig. \ref{fig:resolution_comparison} considering only the high resolution simulation data in orange. Each line represents the median distribution of the reference emulator, black, or the test emulator, colored in shades of blue based on the size of the training set.
  \label{fig:emulator_test}
  }
\end{figure}

Figure \ref{fig:emulator_test} shows the median of the inferred values of all emulators, colored by the training data size. All medians do not deviate more than 0.1 dex between each other, 0.05 dex in some cases, confirming that using all the available data to train the NN does not affect the results.
Moreover, the primary and secondary dependencies on the properties of the IC and the inferred value of $M_1/M_2$ and $R_{1/2}$ are consistent between all emulators, corroborating that our results do not critically depend on the training set of the NN.

\FloatBarrier
\section{Projecting 3D properties} \label{sec:app:projections}

\begin{figure}[h!]
\centering
\includegraphics[width=0.45\textwidth]{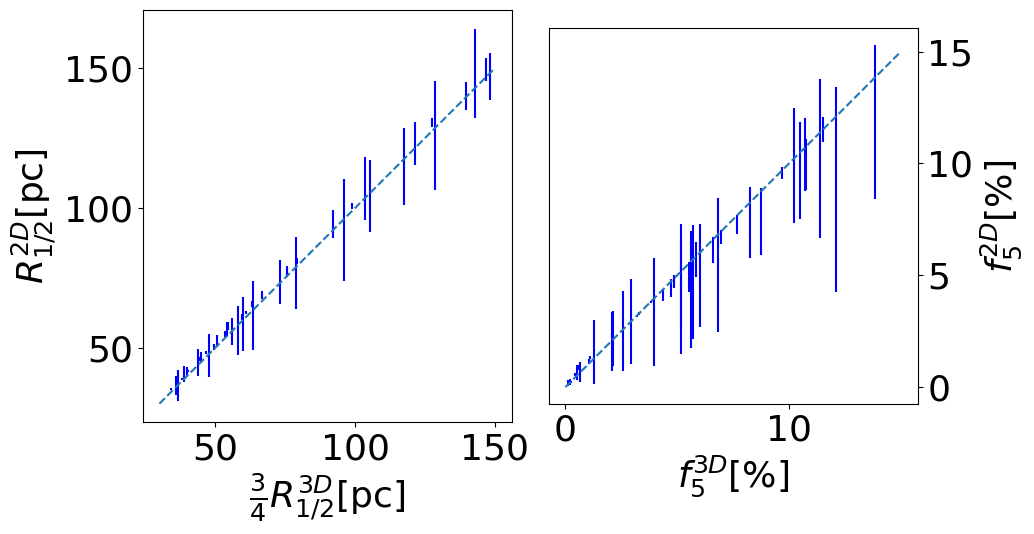}
  \caption{\commenta{Distribution of the half-mass radius (left) and $f_5$ (right) of the post-merger galaxy on the 3D-2D plane. 
  Each solid line represents a simulation in the suite, spanning from the 16\%-th to the 84\%-th percentiles.
  The dashed lines in the two panels are the plane bisectors. The 3D half-mass value is multiplied by 3/4 to account for the projection, as reported by \cite{wolf:2010}. }
  \label{fig:projection_test}
  }
\end{figure}

\commenta{Since the merger orbital plane is fixed in the suite, projecting both $R_{1/2}$ and $f_5$ on the XY plane can have systematic effects on the measures. To minimize such effects and to account for the asymmetry of the post-merger galaxies, we project the galaxy into $8000$ different planes uniformly distributed in the 3D space and measure $R_{1/2}$ and $f_5$ for each projection. We take the median and 16\%-th and 84\%-th percentiles of the resulting distributions as the value and errors, respectively.  
The results of the ensemble average are shown in Fig. \ref{fig:projection_test}, where each solid line represents a simulation and spans the interval between the 16\%-th and the 84\%-th percentile, and the dashed line is the plane's bisector. As reported by \cite{wolf:2010}, the ratio between the 2D and 3D half-mass radii can be approximated to 3/4 for many profiles, which is also true in our post-merger simulations. The 3D values of $f_5$ are within the error bars of the projected $f_5$, suggesting that the parameter is negligibly affected by the projection procedure.}

\end{appendix}
\end{document}